\pdfoutput=1 
\documentclass[dvipsnames]{elsarticle}
\usepackage{hyperref}
\usepackage[usenames]{color}
\usepackage[utf8]{inputenc}
\usepackage[T1]{fontenc}
\usepackage{siunitx}
\usepackage{subcaption}

\usepackage{tikz}
\usepackage{environ}
\makeatletter
    \newsavebox{\measure@tikzpicture}
        \NewEnviron{scaletikzpicturetowidth}[1]{%
            \def\tikz@width{#1}%
            \def\tikzscale{1}\begin{lrbox}{\measure@tikzpicture}%
            \BODY
            \end{lrbox}%
            \pgfmathparse{#1/\wd\measure@tikzpicture}%
            \edef\tikzscale{\pgfmathresult}%
            \BODY
        }
\makeatother
\usetikzlibrary{math}
\usetikzlibrary{calc}
\usetikzlibrary{decorations.pathreplacing, decorations.markings}
\usetikzlibrary{patterns, arrows, arrows.meta}
\usepackage{xcolor}
\definecolor{WP3_Orange}{HTML}{E69F00}
\definecolor{WP3_Vermillion}{HTML}{D55E00}
\definecolor{WP3_Bluish_green}{HTML}{009E73}
\definecolor{WP3_Blue}{HTML}{0072B2}

\DeclareSIUnit{\ele}{\mbox{$\mathrm{e}^{\text{-}}$}}
\DeclareSIUnit[number-unit-product = ]{\percent}{\%}
\DeclareSIUnit{\raddose}{rad}
\DeclareSIUnit{\tid}{\mbox{\si{\kilo\raddose}}}
\DeclareSIUnit{\niel}{\mbox{\SI{1}{\mega\electronvolt}\ n$_{\mathrm{eq}}$\ \si{\per\cm\squared}}}
\DeclareSIUnit{\pixel}{pixel}

\newcommand{\Vh}{V_{\mathrm{pulseh}}}
\newcommand{\Vl}{V_{\mathrm{pulsel}}}
\newcommand{\Vb}{V_{\mathrm{casb}}}
\newcommand{\Vn}{V_{\mathrm{casn}}}

\newcommand{\Vs}{V_{\mathrm{shift}}}
\newcommand{\Ir}{I_{\mathrm{reset}}}
\newcommand{\Id}{I_{\mathrm{db}}}
\newcommand{\Ib}{I_{\mathrm{bias}}}
\newcommand{\Ibn}{I_{\mathrm{biasn}}}
\newcommand{\Iref}{I_{\mathrm{ref}}}
\newcommand{\Vref}{V_{\mathrm{ref}}}
\newcommand{\Vrc}{V_\mathrm{rcas}}
\newcommand{\Cinj}{C_{\textrm{inj}}}

\newcommand{\sifl}{Si-K$_{\alpha}$ }
\newcommand{\siesc}{Mn-K$_{\alpha, \beta}$ -- Si-K$_{\alpha, \beta}$}
\newcommand{\mnka}{Mn-K$_{\alpha}$ }
\newcommand{\mnkb}{Mn-K$_{\beta}$ }

\journal{Nucl.\ Instrum.\ Methods Phys.\ Res.\ A}
\begin{document}%
\begin{frontmatter}
\title{Characterisation of the first wafer-scale prototype for the ALICE ITS3 upgrade: the monolithic stitched sensor (MOSS)}

\author[1]{Omar Abdelrahman}
\author[2]{Gianluca Aglieri Rinella}
\author[3,4]{Luca Aglietta}
\author[3,4]{Giacomo Alocco}
\author[5,6]{Matias Antonelli}
\author[6]{Roberto Baccomi}
\author[7,8]{Francesco Barile}
\author[9]{Pascal Becht}
\author[4]{Franco Benotto}
\author[3,4]{Stefania Maria Beolè}
% ...

\affiliation[1]{organization={Technische Universität München}, city={Munich}, country={Germany}}
\affiliation[2]{organization={European Organisation for Nuclear Research (CERN)}, city={Meyrin}, country={Switzerland}}
\affiliation[3]{organization={Universit\`a degli studi di Torino}, city={Torino}, country={Italy}}
\affiliation[4]{organization={Istituto Nazionale di Fisica Nucleare (INFN), Sezione di Torino}, city={Torino}, country={Italy}}
\affiliation[5]{organization={Universit\`a degli studi di Trieste}, city={Trieste}, country={Italy}}
\affiliation[6]{organization={Istituto Nazionale di Fisica Nucleare (INFN), Sezione di Trieste}, city={Trieste}, country={Italy}}
\affiliation[7]{organization={Dipartimento Interateneo di Fisica ‘M. Merlin’}, city={Bari}, country={Italy}}
\affiliation[8]{organization={Istituto Nazionale di Fisica Nucleare (INFN), Sezione di Bari}, city={Bari}, country={Italy}}
\affiliation[9]{organization={Ruprecht Karls Universit\"at Heidelberg}, city={Heidelberg}, country={Germany}}
% ...

\cortext[cor1]{Corresponding author}

\begin{abstract}

This paper presents the characterisation and testing of the first wafer-scale monolithic stitched sensor (MOSS) prototype developed for the ALICE ITS3 upgrade that is to be installed during the LHC Long Shutdown~3 (2026--2030). The MOSS chip design is driven by the truly cylindrical detector geometry that imposes that each layer is built out of two wafer-sized, bent silicon chips. The stitching technique is employed to fabricate sensors with dimensions of \qtyproduct{1.4 x 25.9}{\cm}, thinned to \SI{50}{\micro\metre}.
The chip architecture, the in-pixel front-end, the laboratory and in-beam characterisation, the susceptibility to single-event effects, and the series testing are discussed. The testing campaign validates the design of a wafer-scale stitched sensor and the performance of the pixel matrix to be within the ITS3 requirements.
The MOSS chip demonstrates the feasibility of the ITS3 detector concept and provides insights for further optimisation and development.
\end{abstract}


\end{frontmatter}

\tikzset{
	beamarrow/.style={
		decoration={
			markings,mark=at position 1 with 
			{\arrow[scale=2,>=stealth]{>}}
		},postaction={decorate}
	}
}
\tikzset{
	pics/.cd,
	vector out/.style={
		code={
		\draw[#1, thick] (0,0)  circle (0.15) (45:0.15) -- (225:0.15) (135:0.15) -- (315:0.15);
		}
	}
}
\tikzset{
	pics/.cd,
	vector in/.style={
		code={
		\draw[#1, thick] (0,0)  circle (0.15);
		 \fill[#1] (0,0)  circle (.05);
		 }
	}
}
\tikzset{
	global scale/.style={
		scale=#1,
		every node/.style={scale=#1}
	}
}
\def\centerarc[#1] (#2)(#3:#4:#5) 
	 { \draw[#1] ($(#2)+({#5*cos(#3)},{#5*sin(#3)})$) arc (#3:#4:#5); }

\section{Introduction}
\label{sec:intro}

Monolithic Active Pixel Sensors (MAPS) are used in high-energy physics experiments as they enable the construction of ultra-thin, large-scale detectors.
They were first used in a collider environment in the STAR pixel detector~\cite{star}. The ALPIDE sensor implemented in a \SI{180}{\nm} CMOS imaging process~\cite{ALPIDE-proceedings-1, ALPIDE-proceedings-2, ALPIDE-proceedings-3} was used on a much larger scale in the Inner Tracking System 2 (ITS2) of the ALICE experiment at the CERN Large Hadron Collider (LHC)~\cite{ls2paper}.
Since 2020, the ALICE Collaboration~\cite{alice-collaboration}, in synergy with CERN EP R\&D~\cite{eprnd}, has been carrying out extensive R\&D for its ITS3 upgrade~\cite{ITS3_TDR}, which will replace the three inner layers of the ITS2 and which is scheduled for installation during the LHC Long Shutdown~3 (LS3, 2026--2030). The main objective of the ITS3 detector is to reduce the material budget from the current \SI{0.36}{\percent}~X$_{0}$ per layer down to about \SI{0.09}{\percent}~X$_{0}$ per layer, coming mostly from the sensors themselves.
Furthermore, a new beam pipe with an inner radius of \SI{16}{\milli\meter} and wall-thickness of \SI{500}{\micro\meter} is foreseen, which allows the innermost layer to be installed as close as possible to the beam axis.
The required\footnote{\label{fn:radhard}Radiation tolerance requirement has been updated w.r.t.~Ref.~\cite{ITS3_TDR}, while the corresponding measurements presented in this paper have been carried out at 2.5~times higher doses compared to those quoted in Ref.~\cite{ITS3_TDR}.} radiation tolerance of the detector is \SI{4}{\kilo\gray} of Total Ionising Dose (TID) and \SI{4e12}{\niel} of Non-Ionising Energy Loss (NIEL).

To meet these constraints, the ALICE Collaboration has adopted the \SI{65}{\nm} CMOS imaging technology developed by Tower Partners Semiconductor Co.~\cite{tower}. This technology was validated for ALICE applications and beyond with a set of test structures~\cite{apts_paper,dpts_paper,dpts_2, apts_opamp_paper}.
The \SI{300}{\mm} wafer size and the stitching technique~\cite{tower-stitching, Leitao_MOSS} allow the fabrication of sensors substantially larger than the typical reticle size (about \qtyproduct{3x2}{\centi\meter}), with sensor dimensions reaching \qtyproduct{10x27}{\centi\meter}. Thinning to \SI{50}{\micro\meter} thickness allows these wafer-scale sensors to be bent and form a self-supporting, truly cylindrical structure. Dedicated studies have demonstrated that MAPS maintain their performance after bending~\cite{bent_1,bent_2}.
Due to a relatively low power consumption of about \SI{40}{\mW\per\cm^2}, air cooling becomes feasible, thereby minimising the need for complex cooling and mechanical support structures.

The ALICE ITS3 detector will consist of two truly cylindrical half-barrels.  Each half-barrel consists of three sensor layers with a length of \SI{26.6}{\cm} at radii of \SIlist{19.0; 25.2; 31.5}{\milli\meter}~\cite{ITS3_TDR}. Each layer of a half-barrel is formed from a single silicon sensor, built as an array of repeated smaller layout components described in the next chapter. The sensor is physically bent into a half-cylindrical shape and supported by ultra-light carbon foam structures. Electrical interconnection is provided solely through wire-bonding at the ends of the half-cylinder.
The production of such large-area pixel sensors is a novel development in the field of high-energy physics experiments, and the prototype sensor discussed in this article aims to assess the feasibility and performance characteristics of this technology and concept.

The MOnolithic Stitched Sensor (MOSS) was fabricated in 2023 as part of the Engineering Run~1 (ER1). 
MOSS measures \qtyproduct{1.4x25.9}{\cm} and served as a demonstrator for the stitching process and the performance of the pixel matrix under ITS3 operating conditions.
The development goals of the MOSS design included:
(a) gaining experience with the stitching technique to design large sensors that meet the integration requirements of ITS3;
(b) studying topologies for distributing power and signals using metal interconnects that span wafer-scale distances;
(c) investigating yield and constraints related to Design for Manufacturability (DfM) rules;
(d) evaluating the performance of large-area pixel arrays; and
(e) analysing noise, power consumption, leakage, and variability in electrical characteristics across very large sensors.
A comprehensive characterisation campaign was carried out on the MOSS sensor to gain knowledge and to assess its compliance with ITS3 requirements~\cite{ITS3_TDR}. The MOSS sensor and the findings from this campaign are presented and discussed in this article.

\section{MOSS sensor}
\label{sec:moss_sensor}
\label{subsec:architecture}

\begin{figure}[!htb]
    \centering
    \includegraphics[width=\linewidth]{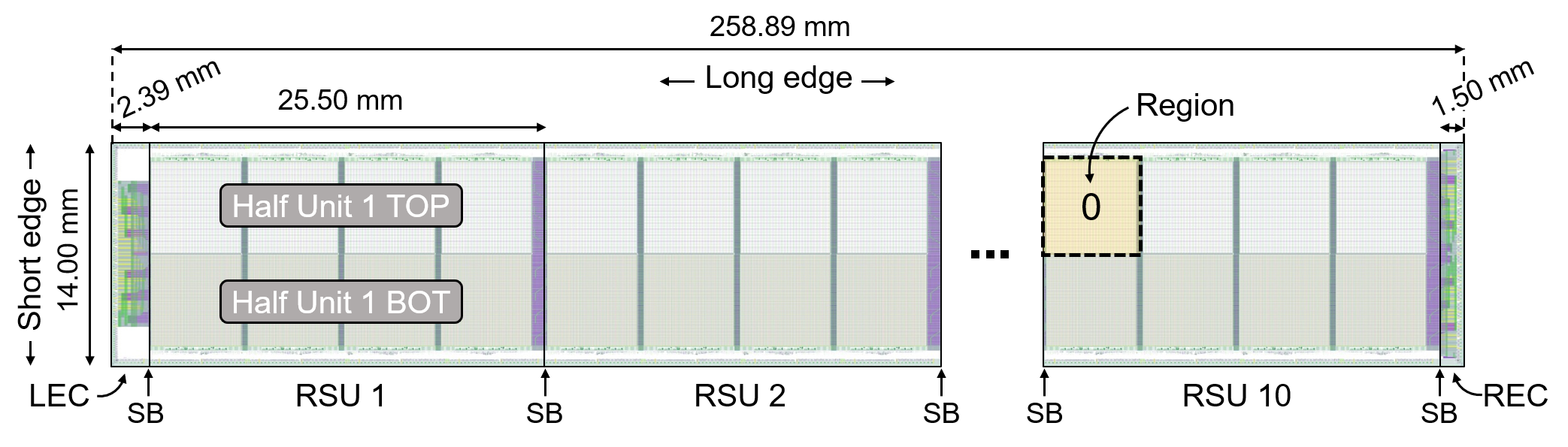}
    \caption{MOnolithic Stitched Sensor (MOSS) layout. Left End-Cap (LEC), Repeated Sensor Unit (RSU), Right End-Cap (REC), and Stitching Boundaries (SB) are indicated.}
    \label{fig:moss_architecture}
\end{figure}

The MOSS prototype sensor, exploring the application of stitching for the fabrication of wafer-scale MAPS, is schematically shown in Fig.~\ref{fig:moss_architecture}.
The design is made of three layout components in the design reticle: the Left End-Cap (LEC), the Right End-Cap (REC), and the Repeated Sensor Unit (RSU). The full sensor consists of a linear array of ten abutting RSUs, completed by one LEC and one REC at the respective ends, resulting in a one-dimensional stitched assembly.
Each RSU is subdivided into two symmetrical sections referred to as \textit{top and bottom half-units}. 
Every half-unit contains four \textit{regions}, each containing one pixel matrix and the related biasing, control and readout.
The matrices of top regions have $256\times256$ square pixels with a pitch of \SI{22.5}{\micro\metre}. The ones of bottom regions have $320\times320$ square pixels with a pitch of \SI{18.0}{\micro\metre}.
The area occupied by pixels within each RSU is about \SI{265}{\milli\meter\squared} out of the total RSU area of \SI{357}{\milli\meter\squared}.
In total, the MOSS sensor comprises approximately 6.72~million~pixels. The large and small pixel pitches in the top and bottom halves of the sensor allow testing of low and high integration densities, respectively. This feature was intended to evaluate how integration density may affect the functional yield of a large-area stitched sensor.

\begin{figure}[!htb]
    \centering
    \includegraphics[width=0.4\linewidth,trim=1cm 1cm 1cm 1cm, clip, keepaspectratio]{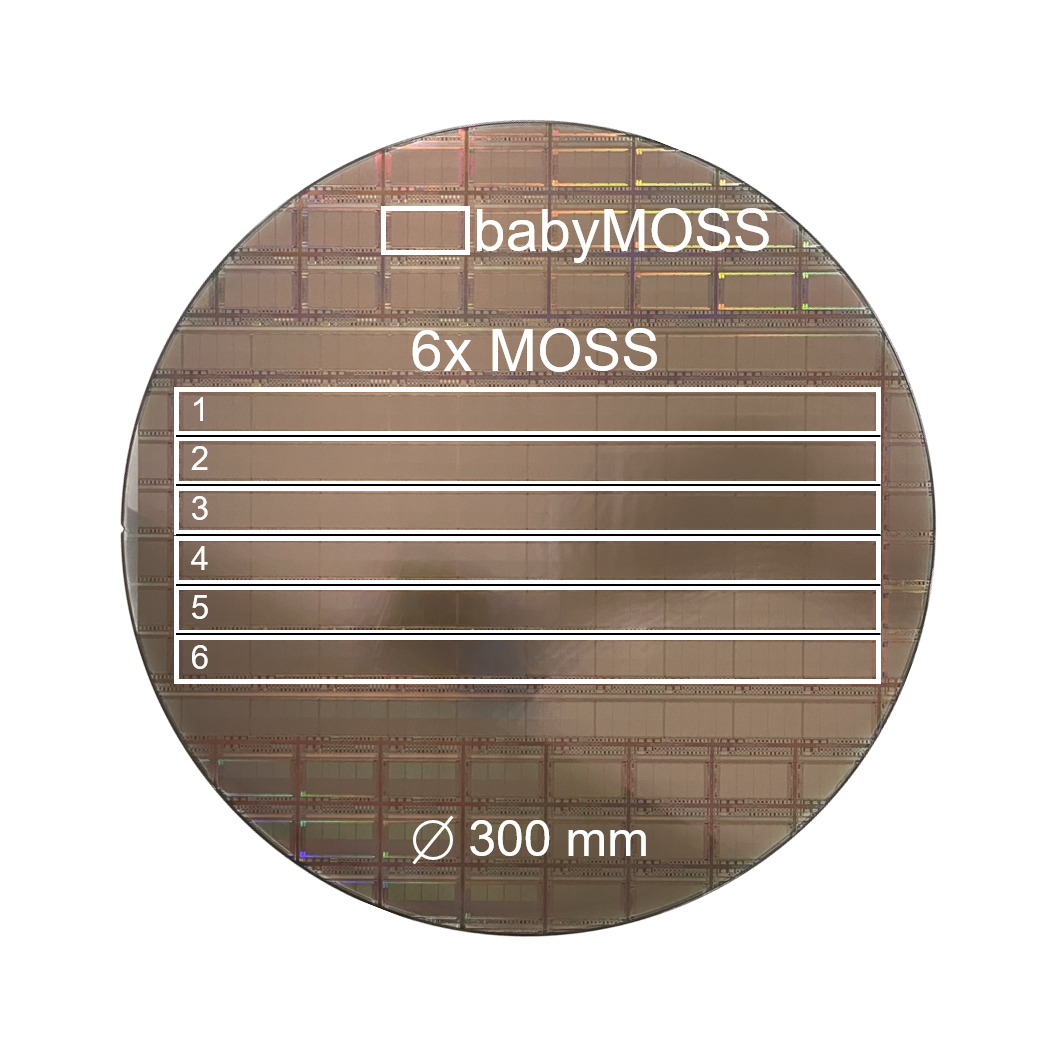}
    \caption{Processed wafer with 6 numbered MOSS sensors in the centre, and one of the overall 23 babyMOSS sensors labelled near the top of the wafer.}
    \label{fig:ER1_processed_MOSS}
\end{figure}

One 300-mm-diameter wafer comprises six MOSS sensors as shown in Fig.~\ref{fig:ER1_processed_MOSS}. The physical area covered by six MOSS sensors is similar to the area of the outermost ITS3 layers, while five and four adjacent MOSS sensors cover surfaces that are similar to the ones of the middle and innermost layers, respectively. Additionally, each wafer contains 23 \textit{babyMOSS} structures, comprising one RSU, LEC, and REC each. Although smaller than the MOSS, these are effectively fully functional sensors with only one RSU.

\begin{figure}[!htb]
    \centering
    \includegraphics[width=\linewidth]{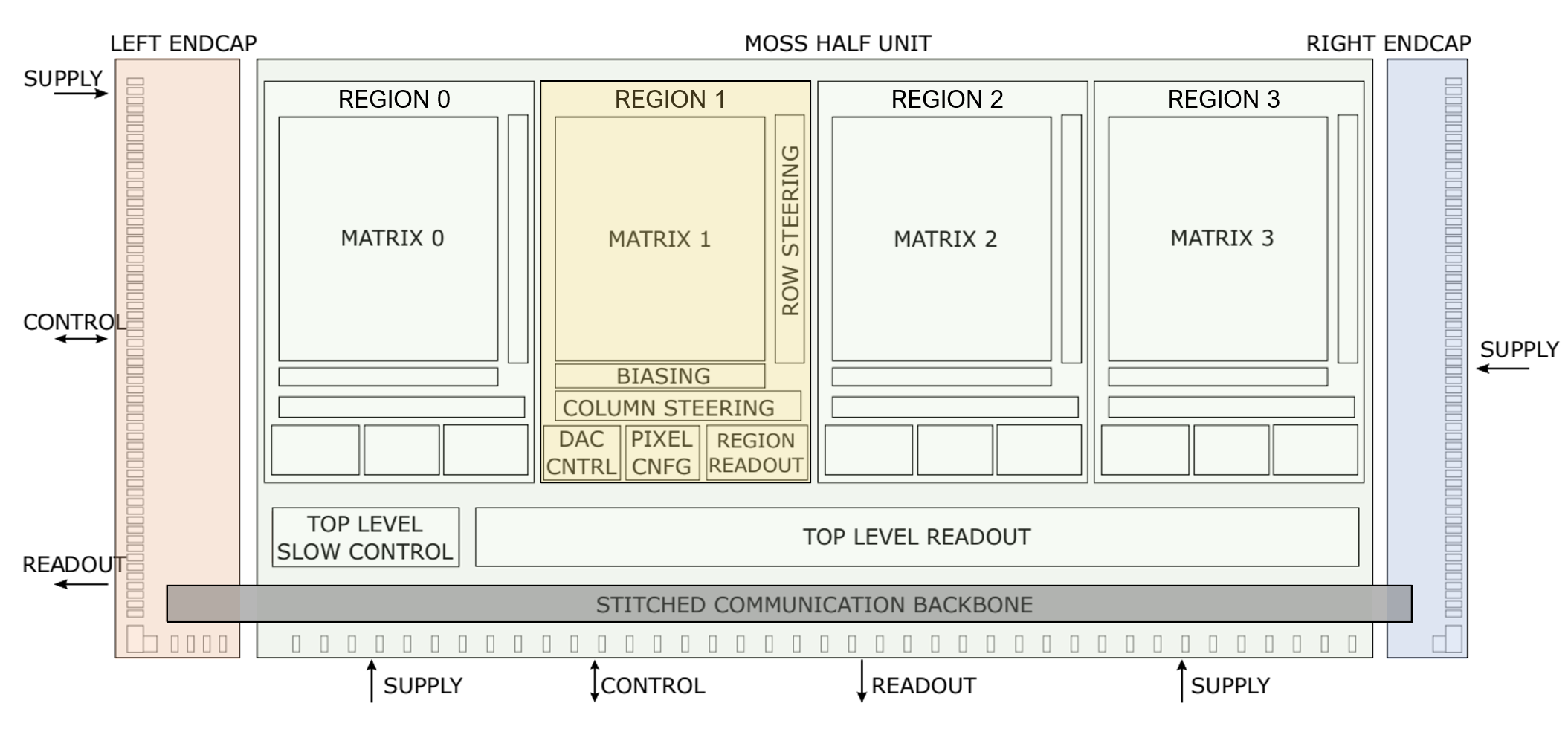}
    \caption{MOSS sensor block diagram with one bottom half-unit of a RSU, a LEC, and a REC. Supply, control, and readout lines are schematically indicated. Functional blocks within the half-unit are labelled. The stitched communication backbone spans the full length of the sensor, crossing the stitching boundaries. Region~1 is highlighted, illustrating the contained blocks.}
    \label{fig:moss_block_diagram}
\end{figure}

The functional block diagram of the MOSS sensor is provided in more detail in Fig.~\ref{fig:moss_block_diagram}, illustrating a bottom half-unit of a RSU, a LEC, and a REC. 
Each of the 20 half-units (10~top, 10~bottom) can also function autonomously via separate wire bond pads situated along the sensor’s long edge, enabling individual operation and characterisation. This modular design of the first stitched prototype enables the distinction between potential faults at the individual half-unit level and those originating from the stitching process. Individual half-units can be independently powered, isolated, and tested even if others malfunction. Signal and power routing between RSUs and the end-caps is achieved by metal-wiring crossing the RSU boundaries, which is produced by stitching. The communication lines, referred to as the stitched communication backbone, are routed over the periphery outside the regions and close to the wire bond pads, as illustrated in Fig.~\ref{fig:moss_block_diagram}. The powering lines are routed as a grid across the entire sensor area and above the pixel arrays.

Two modes of operation are supported: one provides independent control and readout for each half-unit via the bond pads along the sensor's long edge, while the other enables control and readout through the I/Os in the LEC. Power is supplied via the top and bottom bond pads for all RSUs. Because of the fine subdivision of supply nets, the limitations of the MOSS-sensor’s metal stack in terms of conductivity, and to prevent excessive voltage drops along the sensor length, power can be supplied via the LEC and REC only to the leftmost (RSU~1) and rightmost (RSU~10), respectively. Each half-unit comprises individual analogue (AVDD, AVSS) and digital (IOVDD, DVDD, DVSS) power nets\footnote{The convention of *VDD and *VSS suffixes was chosen, representing the positive supply and corresponding ground, respectively.}. The IOVDD net supplies the level-shifting circuitry that translates on-sensor \SI{1.2}{\volt} to off-sensor \SI{1.8}{\volt} signal levels. Separate global power nets (BBVDD, BBVSS) are available for powering the stitched backbone circuitry, with one net for the top half and a separate one for the bottom half of the sensor.
On the LEC, the control and readout I/Os of the top and bottom backbones have dedicated supply nets (BBIOVDD).
The sensor substrate biasing net (PSUB) is global for the entire sensor. It is used to reverse bias the charge-collection diodes.
Each half-unit has an additional, multiplexed analogue I/O pad on the long edge, used for monitoring and characterizing the on-sensor DACs of the four regions within the half-unit.

\subsection{In-pixel front-end}

\label{sec:frontend}

\begin{figure}[!htb]
    \centering
    \includegraphics[width=0.95\linewidth]{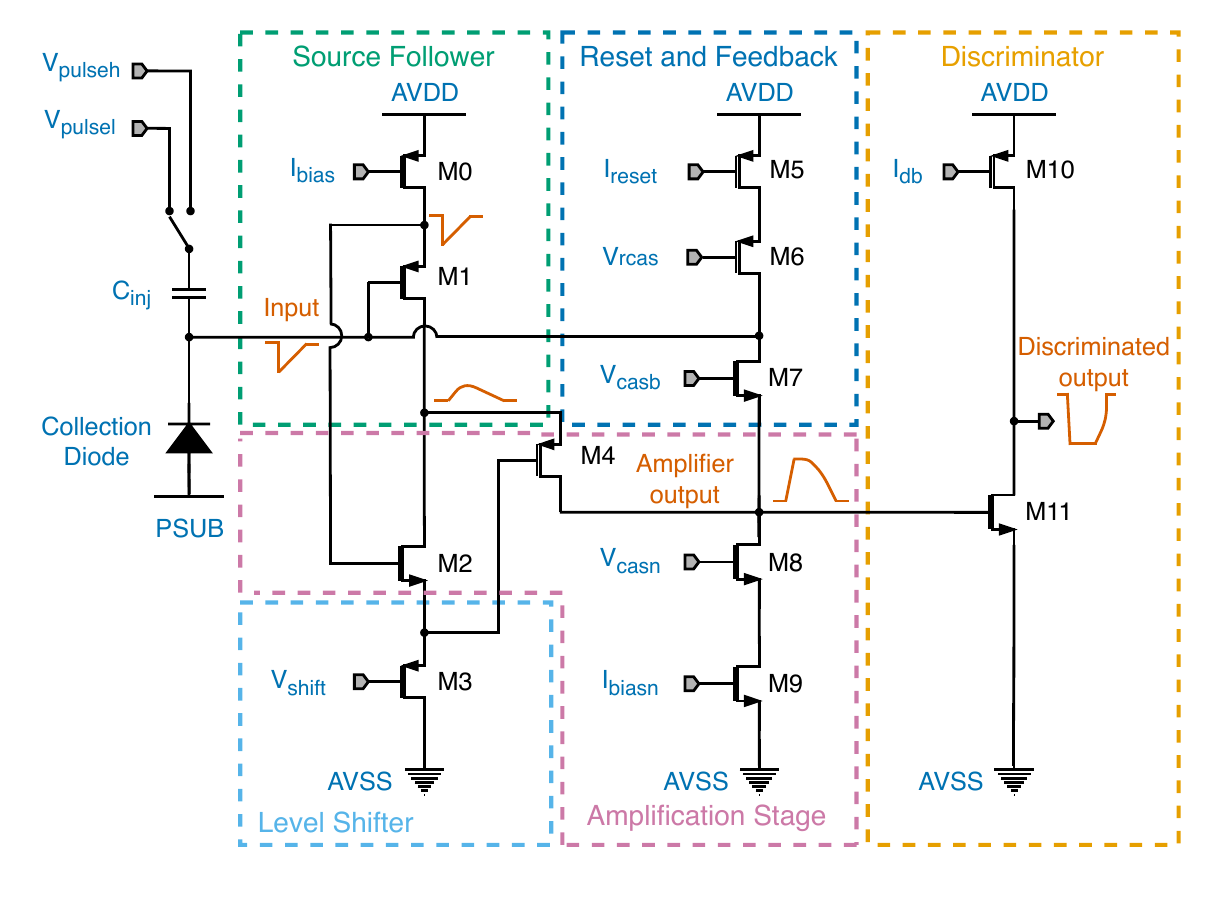}
    \caption{Simplified front-end schematic. Control voltages are applied to the gates of the corresponding transistors, while bias currents are provided via current mirrors (indicated as transistors with a current supplied to the gate, e.g.~M0). The orange traces illustrate the characteristic voltage signals at key nodes within the front-end circuit.}
    \label{fig:frontend}
\end{figure}

The in-pixel front end is designed to bias the collection electrode, it amplifies the charge signal and applies a threshold to it to determine whether the pixel was hit or not~\cite{FE}. The front-end is not a charge amplifier in the classical sense: it  profits from the low capacitance at the input node, and hence the relatively large voltage excursion caused by the collected signal charge. The amplification is carried out both by M1 and M2, who each contribute to the signal of the amplifier output. Its operating point is defined by four currents $\Ib$, $\Ibn$, $\Ir$, and $\Id$, and four voltages $\Vb$, $\Vn$, $\Vs$, and $\Vrc$, indicated in the simplified schematic in Fig.~\ref{fig:frontend}.
To meet the stringent power consumption constraints, most transistors operate in weak inversion and with very low biasing currents\footnote{The precise values of biasing currents and voltages are discussed in Sec.~\ref{subsubsec:two_params_scan}.}. A brief overview of the operating principle of the in-pixel front-end is outlined in the following.

Charge collected by the pixel diode causes a voltage drop at the input node of the front-end (gate of M1). The input source-follower transistor M1 provides a high input impedance. The source of M1 reproduces the input voltage and is connected to the gate of the main amplifying transistor M2. The M2 amplifier, M4 cascode, and M9 current sink (together with its related cascode M8) form a folded-cascode amplification stage.  M8 stabilises the $\Ibn$ current-sink branch, and its gate voltage $\Vn$ is set based on the operating conditions of M7 and M9.
M0 provides $\Ib$, the main current of the front-end. Increasing $\Ib$ increases the gain and decreases equivalent input noise and response time, at the cost of increased power consumption. $\Ibn$ is set significantly lower than $\Ib$, nominally 1/10 of $\Ib$, to boost the output impedance of the amplifier and therefore its gain.

The level-shifting transistor M3, steered by $\Vs$, increases the bias on the collection diode thus reducing its capacitance, up to the limit where M0 is pushed out of saturation, i.e.~the main front-end biasing current is reduced.
Transistors M5, M6, and M7 constitute the feedback to the input node that sets the input voltage. 
M5 and M6 together form a cascoded current source. This topology provides a better control of the small $\Ir$ current over large matrices. The M6 gate voltage $\Vrc$ is derived from $\Ir$ in the biasing unit (see Sec.~\ref{sec:moss_matrix_dacs}), and its value is typically close to the rail voltage (AVDD). The $\Ir$ current must be set larger than the sensor leakage current, but it must be limited to avoid additional shot noise and degradation of the amplifier gain. $\Ir$ is therefore in the pico-ampere range. 
The $\Vb$ voltage on the gate of M7 and $\Ir$ establish the baseline voltage at the output of the amplifier. 
Upon charge collection, the rising voltage at the amplifier output reduces the gate-source voltage of M7, turning it off and redirecting the $\Ir$ current to reset the collection diode.
A reverse bias of \SI{-1.2}{\volt} is applied to the sensing diodes via the substrate biasing net PSUB. This voltage is unavoidably applied also to the bulk of the NMOS transistors in the pixels. This reverse substrate/bulk bias for the NMOS transistors in the pixel matrix results in a transistor threshold or VT shift, reducing the current drive capability of the NMOS transistors. This is not a show-stopper, but the effect has to be taken into account in the design of the in-pixel circuitry.

The discrimination of the amplifier output signal is implemented with transistors M10 and M11. In static operation, the $\Id$ current is larger than the standby current in M11, and the discriminator output is thus kept close to the supply voltage. When the amplifier output voltage raises following a hit, the current in M11 increases, and if it exceeds $\Id$, the drain voltage of M10 drops to almost ground, indicating a hit. The charge threshold, which is set using one value for all pixels of the entire half unit of a RSU, therefore depends on the amplifier gain (influenced by $\Ib$ and $\Ir$), its output baseline defined by $\Vb$ and $\Ir$, and the discriminator current $\Id$. 

For testing and calibration, charge can be injected at the circuit input by applying a voltage step to the injection capacitance $\Cinj$. This step is produced by switching the capacitor node between $\Vh$ and $\Vl$, where $\Vl$ corresponds to the potential of the AVSS ground net of the DAC that generates $\Vh$ at the periphery of the matrix (see Sec.~\ref{sec:moss_matrix_dacs}).

The MOSS sensor integrates four variants of the in-pixel front-end in addition to the baseline design, to investigate potential optimization. One variant uses a larger input transistor M1 to reduce random telegraph and $1/f$ noise, another features an enlarged discriminating transistor M11 to reduce threshold dispersion, and one includes a larger amplifying transistor M2 to increase gain, albeit with the trade-off of additional input capacitance. The fourth variant employs a modified layout to study the influence of inter-device parasitic capacitances.

\subsection{Pixel matrix and biasing}
\label{sec:moss_matrix_dacs}

Each pixel in the matrix integrates the previously described analogue section together with a dedicated digital section. The pixels are controlled and read out via a network of digital lines organized into orthogonal buses running along each column and row. These signals are routed and buffered through the ROW STEERING and COLUMN STEERING blocks located at the array periphery, as illustrated in Fig.~\ref{fig:moss_block_diagram}. Configuration functions such as masking, pulsing, and resetting of individual pixels are managed by the PIXEL CNFG block in the peripheral region.

Each pixel includes a readout latch that stores the detection of a hit. A global strobe signal, distributed across all pixels and regions within a half-unit, governs the sampling of the pixel discriminator output into the readout latch. A hit is registered when the strobe signal and the discriminated output of the pixel front-end (see Sec.~\ref{sec:frontend}) are simultaneously asserted. The strobe signal is initiated by a user command through the control interface, with its duration and an additional internal delay being configurable. The hit information remains latched in the pixels until a readout command is issued via the control interface, triggering the readout sequence.

Pixel hit readout is managed by the TOP LEVEL READOUT and REGION READOUT blocks. The positions of hit pixels within the array are sequentially encoded through a two-step process: first, scanning rows that contain at least one hit, and then scanning the hit pixels within each selected row. Row and column positions are determined by priority encoders located at the array periphery. For each hit pixel, its row and column addresses are written to a memory buffer in the region readout periphery, after which the corresponding pixel latch is cleared, allowing the encoders to advance to the next hit. This procedure repeats until all pixel latches are cleared. While the readout of the array progresses, the collected hit addresses are assembled into a data frame by the top-level readout unit and transmitted to the data backbone.

\begin{figure}[!htb]
    \centering
    \includegraphics[width=0.9\linewidth]{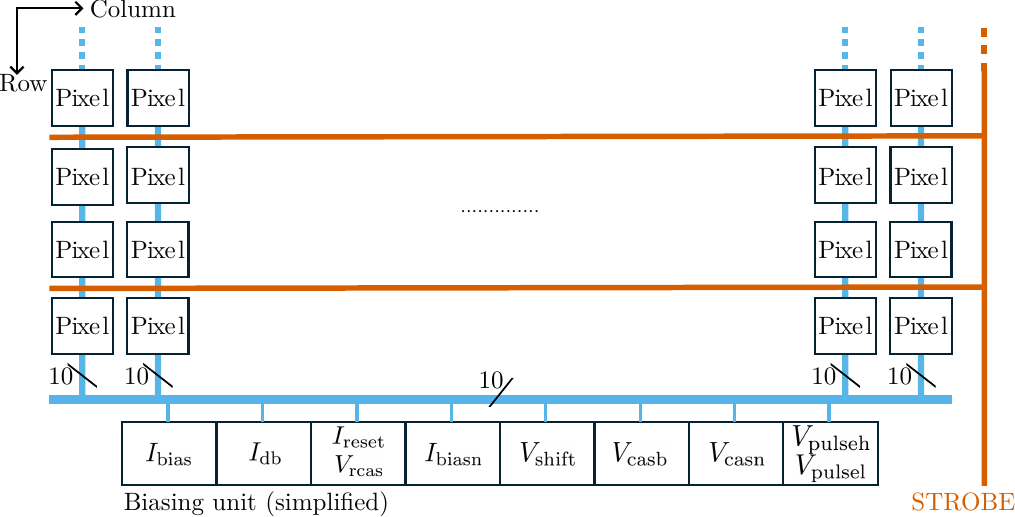}
    \caption{A schematic representation (not to scale) of one region in the bottom half-unit, illustrating the distribution of the tunable analogue bias and the strobe signal to the pixels.}
    \label{fig:layout_strobe_and_biases}
\end{figure}

On the analogue side, all pixels within a region are connected to ten nets that control the biasing of the front-end and the test charge injection circuitry. Four currents, $\Ib$, $\Ibn$, $\Ir$, and $\Id$, together with four voltages, $\Vb$, $\Vn$, $\Vs$, and $\Vh$, are generated by a set of 8-bit DACs in a biasing unit located in the region-specific periphery (see Fig.~\ref{fig:moss_block_diagram}). The voltage $\Vrc$ is derived from the value of $\Ir$, while $\Vl$ is connected to the ground of the $\Vh$ DAC. Each biasing unit includes two internal bandgap circuits that generate reference voltages and currents for the DACs, which can be fine-tuned using dedicated 4-bit DACs.

A scheme of the routing of the biasing and strobe wires is shown in Fig.~\ref{fig:layout_strobe_and_biases}. The digital strobe signal is routed vertically on the side of the pixel matrix and then distributed horizontally across every two pixel rows.
Analogue wires distribute the biasing nets horizontally at the bottom of the matrix and then vertically along each column to the pixels.
The DACs driving each biasing net are distributed horizontally and spaced by \SI{450}{\micro\meter} approximately. Some implications of this layout on the pixel performance are discussed in Sec.~\ref{sec:strobe_and_fhr}.

\subsection{Integration of the pixel sensor within the pixel}
\label{sec:sensordiode}

\begin{figure}[!htb]
    \centering
    \includegraphics[width=0.8\linewidth]{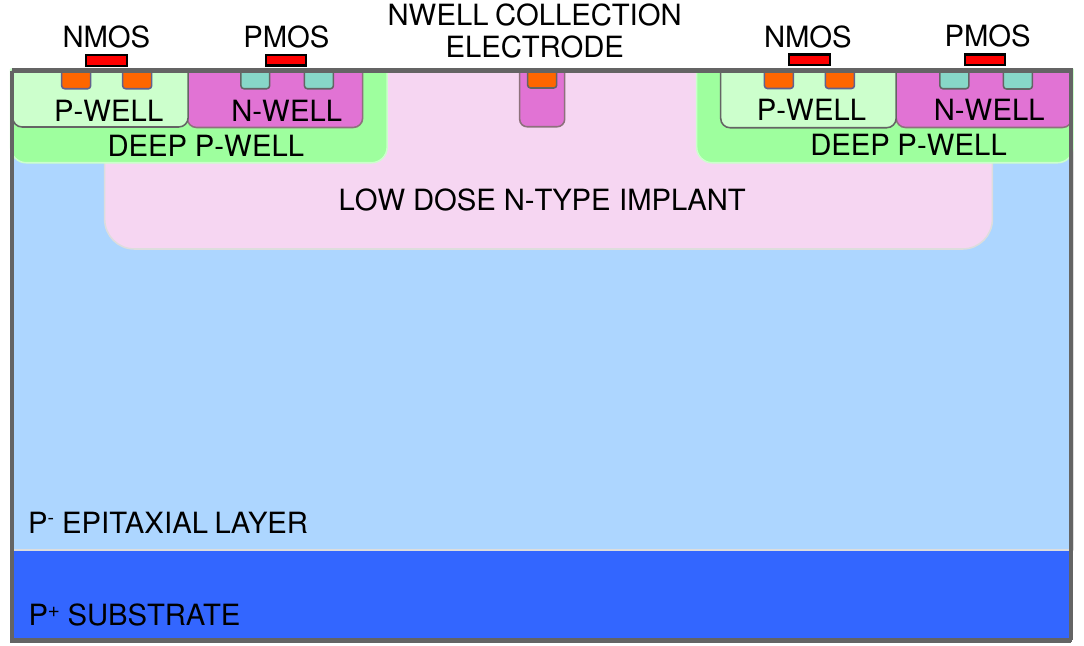}
    \caption{Schematic cross-section of the pixel sensor}
    \label{fig:sensor}
\end{figure}

The design of the pixel sensor shown as a schematic cross-section in Fig.~\ref{fig:sensor} is similar to that of the earlier prototype, the DPTS sensor~\cite{dpts_paper}. It features a low-doped, deep n-type implant in the pixel area which helps to deplete the pixel over its full area. The optimization of the process to implement this pixel and the low-doped, deep n-type implant has been described in \cite{Snoeys:2023hF}.  The gap between the implants of adjacent pixels enhances the lateral field accelerating the charge collection for charges generated near the pixel edges. To investigate the effect of gap size on charge sharing, a layout variation was implemented. In a subset of wafers, and only for pixels with a \SI{22.5}{\micro\meter} pitch, the pixel-to-pixel gap width was increased from the baseline \SI{2.5}{\micro\meter} to \SI{5.0}{\micro\meter}.
\section{Test setup}
\label{sec:testsystem}

\begin{figure}[!hbt]
    \centering
    \includegraphics[width=0.9\linewidth]{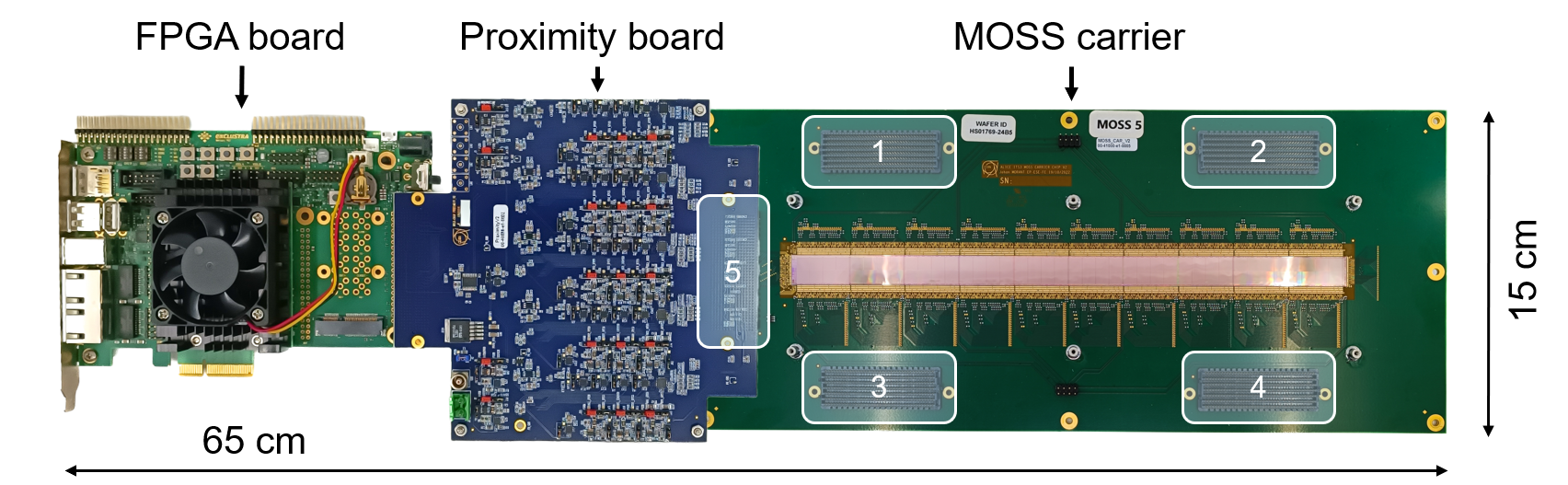}
    \caption{Functional test system. One pair of FPGA board and Proximity board is connected to the carrier connector for the MOSS interfaces on the LEC (5). The additional four connectors \SIrange{1}{4} can connect to FPGA and Proximity boards for powering and characterisation through the long-edge interfaces of the sensor.}
    \label{fig:functional_test_system}
\end{figure}

A dedicated test system, shown in Fig.~\ref{fig:functional_test_system}, has been developed to functionally characterise the MOSS sensor. The sensor is wire bonded onto a passive printed circuit \textit{carrier board}. Custom-designed \textit{Proximity boards} connect to the carrier board, supplying power, and enabling current and voltage monitoring.
Commercial \textit{FPGA boards} are used to operate the Proximity boards and the MOSS sensor.
The sensor slow-control and data-readout lines pass through the Proximity boards and connect directly to the FPGAs.
Communication between the FPGA boards and a PC is established via a USB3 interface. Custom FPGA firmware and Python-based software were developed to steer the setup and the sensor.

The MOSS carrier board has five connectors for FPGA--Proximity board pairs: one is dedicated to the sensor interconnects on the LEC, and four are for the I/Os along the top and bottom long edges of the sensor.
One single FPGA--Proximity board pair is sufficient to test a set of five half-units connected to one of the long-edge connectors. 
The global PSUB voltage is supplied by an external power supply.
A full configuration with five pairs of FPGA--Proximity boards connected to all the carrier connectors is used for series testing, and allows for independent half-unit characterisation, simultaneous operation of all the half-units, and testing via the LEC (see Sec.~\ref{sec:series_testing}).

\subsection{Testbeam setup}
\label{sec:testbeam_setup}

A testbeam telescope was set up to investigate the MOSS in-beam performance in different operating conditions. The system used a single FPGA--Proximity board pair connected to one of the connectors \SIrange{1}{4} in Fig.~\ref{fig:functional_test_system}.
Multiple test campaigns took place at the CERN PS and SPS testbeam facilities from July 2023 to April 2025. 
The results, based on data taken at the CERN PS with a set of representative sensors and a beam of \SI{7}{\giga\electronvolt\per\textit{c}} negative hadrons, are presented in Secs.~\ref{sec:detection_efficiency} and \ref{sec:spatial_resolution}.

\begin{figure}[!htb]
	\centering
    \input{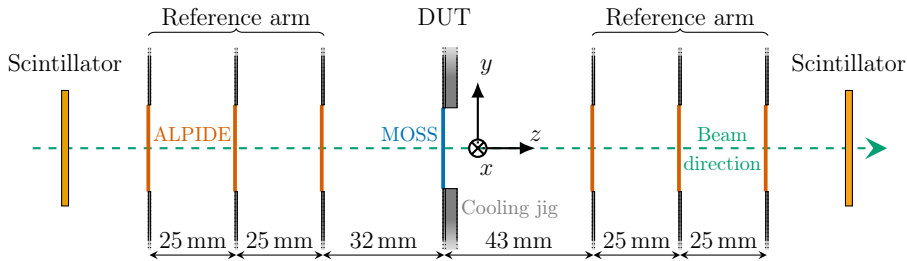}
    \caption{Schematic representation of the setup used for in-beam measurement with ionising particles (not to scale). Six ALPIDE sensors are used as reference planes. A Device Under Test (DUT) is placed between two reference arms. The coincidence of the two scintillator signals is used for triggering.}
    \label{fig:testbeam_setup_sketch}
\end{figure}

Figure~\ref{fig:testbeam_setup_sketch} shows a schematic diagram of the beam telescope. Six reference planes equipped with ALPIDE sensors~\cite{ALPIDE-proceedings-1,ALPIDE-proceedings-2,ALPIDE-proceedings-3} were used to reconstruct particle tracks, with the MOSS sensor placed as a Device Under Test (DUT) in between the reference arms.
An aluminium cooling jig placed on the back of the MOSS carrier was used to keep the sensor at a constant temperature of \SI{27}{\celsius}, corresponding to the operating conditions envisaged for the ITS3~\cite{ITS3_TDR}. The carrier board and the cooling jig featured a cutout corresponding to the location of the pixel matrices in order to limit multiple scattering.
The coincidence of the amplified and discriminated signals of two scintillators triggered the data acquisition.
To reduce the selected events to those containing a single particle track, only trigger signals spaced more than \SI{50}{\micro\second}, with an additional dead time of \SI{100}{\micro\second} after sending the trigger, were accepted.

The EUDAQ~\cite{eudaq} and Corryvreckan~\cite{corryvreckan} frameworks were used for the data acquisition and analysis, respectively. 
Particle tracks were reconstructed by fitting the positions of the clusters found in the reference planes with General Broken Lines~\cite{gbl_formalism}.
A clean data sample was selected by requiring exactly one reconstructed track per event, a single hit on each reference plane, and a good track fit ($\chi^2 < 3$). Clusters recorded by the DUT were associated with tracks if they fell within a circular acceptance window of \SI{30}{\micro\meter} radius, centered on the interpolated intersection of the track with the DUT.

\section{Pixel matrix characterisation}      
\label{sec:pixel_matrix}

The pixel matrix was characterised both in the laboratory and with charged-particle beams, with the aim of measuring detection efficiency, fake-hit rate, and spatial resolution before and after irradiation. BabyMOSS sensors were used interchangeably with MOSS sensors, particularly in measurements where size was a limiting factor, such as irradiation campaigns. Under identical conditions, the performance of the two sensor sizes was indistinguishable within the sensor-to-sensor variations, as expected given that they differ only in the number of Repeated Sensor Units (see Sec.~\ref{subsec:architecture}). The performance of all front-end variants was studied in detail. However, since the differences are minimal and consistent with expectations (see Sec.~\ref{sec:frontend}), only the standard (baseline design) front-end variant will be discussed in the following.

\subsection{Threshold, noise, and fake-hit rate}
\label{sec:thr_noi_fhr}
\label{sec:threshold_noise_fhr}

The effective discrimination threshold applied to the charge-collection diode signal is determined by the in-pixel front-end operating point, primarily by the amplifier output baseline and the discriminator current, adjusted via $\Vb$ and $\Id$, respectively (see Sec.~\ref{sec:frontend}). To measure the charge threshold, injection capacitance $\Cinj$ (see Fig.~\ref{fig:frontend}) is used.
The injected charge is incrementally increased by adjusting $\Vh$, and the pixel output state is recorded for each charge level over multiple repeated injections. The hit probability as a function of the injected charge follows the characteristic S-curve response which can be described by a Gaussian error function~\cite{dpts_paper}. The two parameters of this function, the mean and standard deviation, correspond to the pixel threshold and noise.

\begin{figure}[!htb]
    \centering
    \begin{subfigure}{0.49\linewidth}
        \centering
        \includegraphics[width=\linewidth]{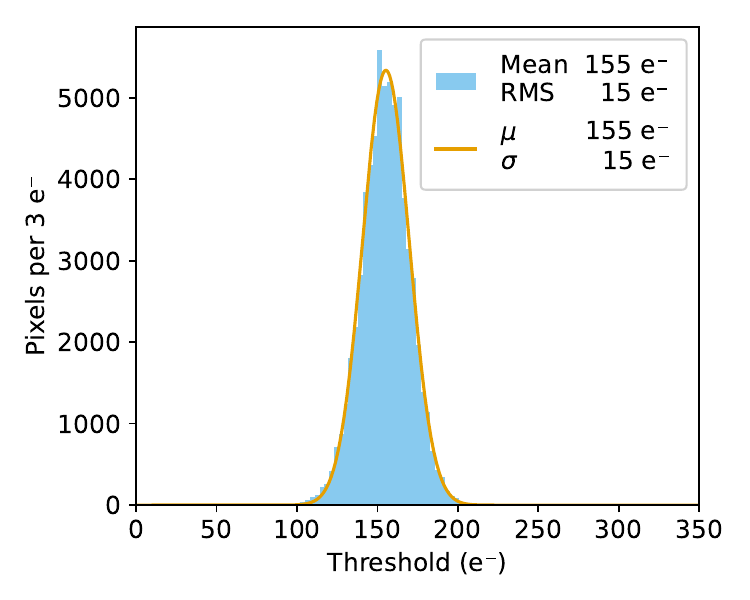}
        \caption{Threshold}
        \label{fig:threshold_dist}
    \end{subfigure}
    \hfill
    \begin{subfigure}{0.49\linewidth}
        \centering
        \includegraphics[width=\linewidth]{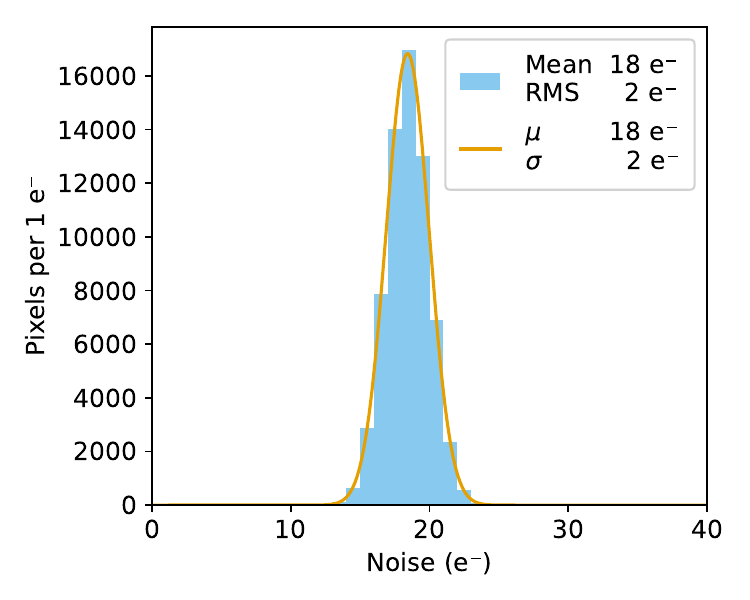}
        \caption{Noise}
    \end{subfigure}
    \caption{Threshold and noise distributions of a reference region (\SI{22.5}{\micro\meter}~pixel pitch, \SI{2.5}{\micro\meter}~gap) under typical operating conditions. Both distributions are well described by a Gaussian fit. The threshold dispersion is comparable to the average pixel noise.}
    \label{fig:threshold_noise}
\end{figure}

Figure~\ref{fig:threshold_noise} shows the distributions of threshold and noise values measured for all pixels in a region under typical operating conditions (see Sec.~\ref{subsubsec:two_params_scan}).
The threshold dispersion, originating from fabrication-induced variations in the in-pixel circuitry, is significant, with an RMS of about \SI{10}{\percent} of the mean threshold value, yet it remains consistent with the earlier prototypes~\cite{dpts_paper}. The measured noise includes contributions from the in-pixel front-end and the sensing node, and its average is comparable to the threshold dispersion.

\begin{figure}[!htb]
    \centering
    \begin{subfigure}{0.49\linewidth}
        \centering
        \includegraphics[width=\linewidth]{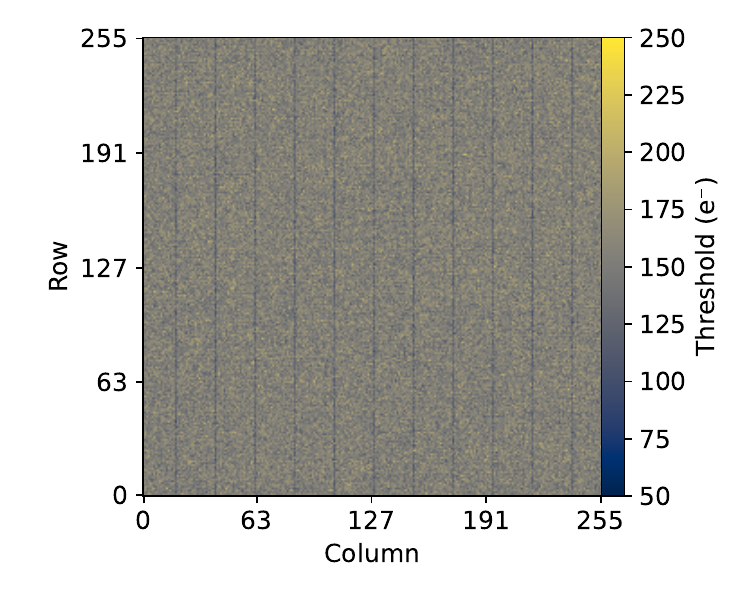}
        \caption{Threshold map}
        \label{fig:thr_map}
    \end{subfigure}
    \hfill
    \begin{subfigure}{0.49\linewidth}
        \centering
        \includegraphics[width=\linewidth]{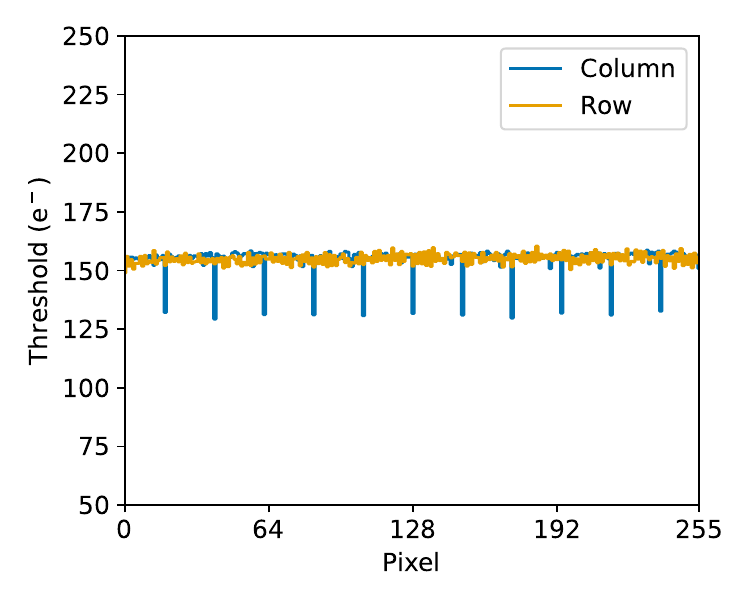}
        \caption{Threshold map profiles}
        \label{fig:thr_profile}
    \end{subfigure}
    \caption{Threshold map of a reference region (\SI{22.5}{\micro\meter}~pixel pitch, \SI{2.5}{\micro\meter}~gap) and its profiles along column and row direction. The threshold dispersion is uniform across the matrix, except of the eleven columns where front-end input is coupled to digital signals steering the charge-injection circuit~\cite{pulsingcircuitry}.}
\end{figure}

The threshold dispersion is mostly uncorrelated with pixel position, as seen in Fig.~\ref{fig:thr_map} where the threshold data is mapped to pixel positions in the matrix. Eleven distinct vertical lines are observed in Fig.~\ref{fig:thr_map} and Fig.~\ref{fig:thr_profile}, and correspond to columns in which the front-end input is coupled to the digital signals steering the charge-injection circuit~\cite{pulsingcircuitry}. Therefore, this is an artefact of the threshold measurement, and is not affecting the sensing performance of the pixels. 

The threshold exhibits a temperature dependence. In measurements with a non-irradiated sensor, a linear decrease of approximately \SI{2}{\ele \per \celsius} was observed over the range \SIrange{20}{35}{\celsius}, consistent with simulations of the in-pixel front-end.

As a result of in-pixel discrimination, noise ultimately manifests as fake hits. The fake-hit rate quantifies how often the sensor registers hits in the absence of particles or charge injection. It accounts for all previously discussed noise sources, including random telegraph noise, as well as systematic effects such as coupling to signal distribution nets (see Sec.~\ref{sec:strobe_and_fhr}). The measurement of the fake-hit rate is performed by issuing several tens of thousands of triggers and recording the number of hits per pixel per unit time.
Pixels registering a fake hit in more than \SI{1}{\percent} of issued triggers are classified as noisy, excluded from the fake-hit rate calculation, and masked from further analysis. In all but a small fraction of cases (see Sec.~\ref{sec:pix_performance}), there are fewer than three such pixels per region, comprising \num{65536} and \num{102400} pixels in top and bottom half-units, respectively. The numbers of pixels that require masking increases after exposing sensors to ionising or non-ionising radiation (see Sec.~\ref{sec:detection_efficiency}).

\subsection{Strobe effect on threshold and fake-hit rate}
\label{sec:strobe_and_fhr}

A parasitic capacitive coupling was identified at the crossings of the analogue-bias and strobe-distribution lines (see Fig.~\ref{fig:layout_strobe_and_biases}). Each crossing introduces a capaci\-tance that was found to cumulatively have a measurable impact on the analogue front-end performance.
Consequently, the falling edge of the strobe signal (see Sec.~\ref{sec:moss_matrix_dacs}, active when low) introduces perturbation to all analogue biases. 
At first, the distributed biasing voltage drops rapidly due to coupling to the strobe signal. The biasing unit then tries to compensate this drop, i.e.~to restore the voltage to its original level. This dynamic response depends on spatial factors, specifically the column number, given by the location of the biasing units (see Fig.~\ref{fig:layout_strobe_and_biases}). The driving strength varies with the distance from the bias origin due to resistance and parasitic capacitance along the path, leading to differences in the amplitude and duration of the resulting perturbations.
A perturbation of a bias connected to the front-end input node, i.e.~$\Vs$ and $\Ib$ biases (see Sec.~\ref{sec:frontend}), produces a signature on the input line similar to that of an injected charge.
As a result, the amplifier output resembles that of a small injected charge, with a typical front-end peak time $\mathcal{O}(\SI{1}{\micro\second})$ and a recovery time of approximately $\SI{10}{\micro\second}$, driven by the recovery time of the biases.

\begin{figure}[htb!]
    \centering
        \includegraphics[width=0.5\linewidth]{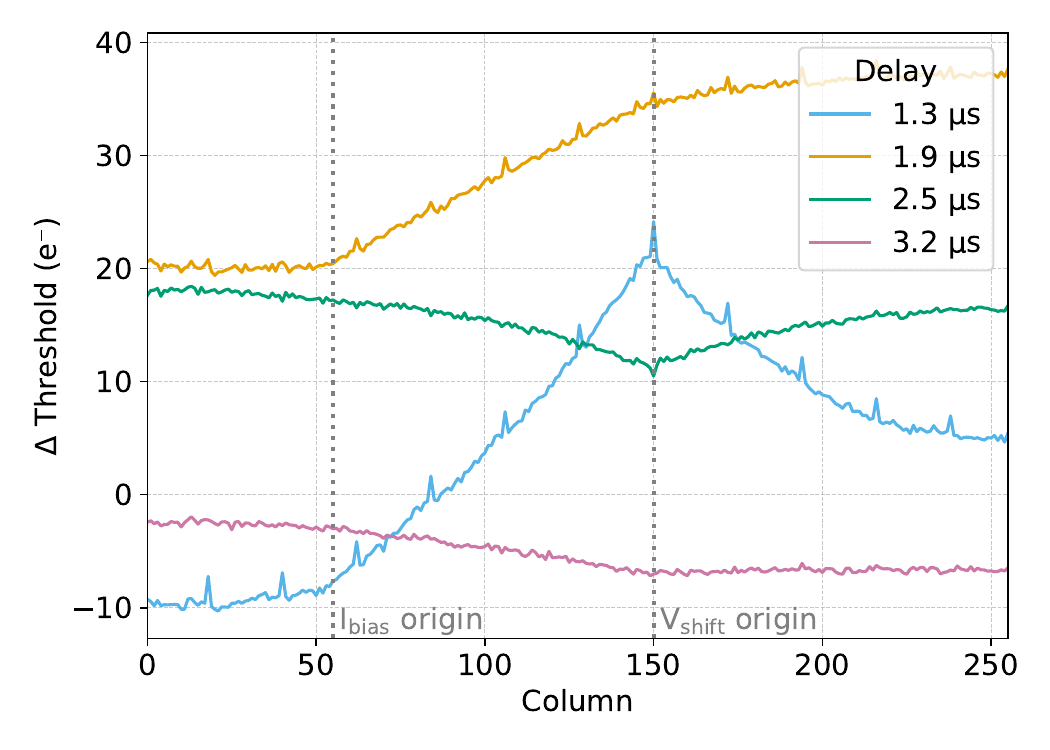}
        \caption{
        Column profile of threshold variation measured at different time delays after the strobe is asserted.
        To isolate the effect of the perturbation, the threshold measured at a delay of \SI{8.8}{\micro\second} when the system is assumed to have recovered from the perturbation, is subtracted from the measurements at shorter delays.
        The small peak substructure arises from the different coupling with the steering signals discussed in Sec.~\ref{sec:threshold_noise_fhr}. }
        \label{fig:perturbation_projection}
\end{figure}

The effect of the perturbation on the threshold is visible in Fig.~\ref{fig:perturbation_projection}, showing the average threshold variation as a function of column, at different delays between the strobe signal and the charge injection. To isolate the effect of the perturbation, the threshold measured at a delay of \SI{8.8}{\micro\second} when the system is assumed to have recovered from the perturbation, is subtracted from the measurements at shorter delays.

The profiles, especially looking at short delays (e.g.~$\SI{1.3}{\micro \second}$), reveal characteristic symmetries at the columns associated with the biasing unit connection to the matrix of $\Vs$ (column 150) and $\Ib$ (column 55).
Furthermore, the threshold shows different patterns at different delays, reflecting the temporal evolution of the perturbation. The difference to the reference is largest at short delays (close to the start of the strobe) and smallest at large delays indicating the recovery from the perturbation. The reference threshold at \SI{8.8}{\micro\second} is chosen because no remaining column dependence or time dependence is observed beyond this point, which also agrees with the expected bias recovery time from simulation.

\begin{figure}[!htb]
        \centering
        \includegraphics[width=0.5\linewidth]{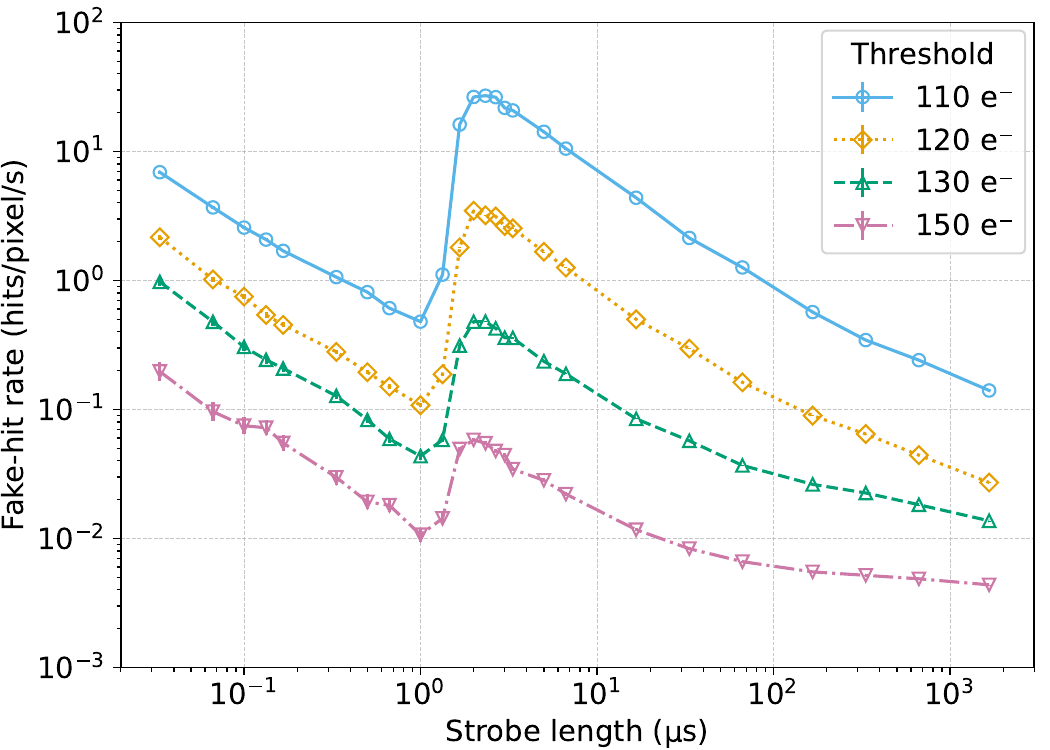}
        \caption{Fake-hit rate as a function of strobe length at different thresholds. The strobe perturbation causes a sharp increase in the fake-hit rate at a strobe length of about $\SI{2}{\micro\second}$. The apparent rise at very short strobe lengths and the apparent decrease at very long strobe lengths are misleading due to the construction of the fake-hit rate observable.
        }
        \label{fig:fhr_vs_strobe}
\end{figure}

Figure~\ref{fig:fhr_vs_strobe} shows the dependence of the fake-hit rate on the strobe signal, illustrating the cumulative effect of perturbations over time on the probability of observing a fake hit. For random noise, the number of fake hits per unit time is expected to be independent of strobe length. In practice, however, two systematic effects modify the measured fake-hit rate. At very short strobe lengths, the measurement is dominated by the few noisiest, continuously active pixels. At very long strobe lengths, the fake-hit rate is underestimated because the in-pixel latch can be asserted only once, limiting the contribution of the noisiest pixels to a single fake hit. Consequently, the dependence of the fake-hit rate on strobe duration is expected to decrease monotonically. Instead, a sudden increase is observed approximately $\SI{2}{\micro\second}$ after the strobe is asserted. This behaviour is consistent with the timescale for amplifying a small charge (noise injection) at the input when the strobe begins.

This finding leads to the conclusion to operate the sensor during the pixel matrix characterisation with a short strobe length of $\SI{0.6}{\micro \second}$. It mitigates the impact of the perturbation by closing the strobe window before the perturbation has a significant effect on the fake-hit rate and ensures that the fake-hit rate is not underestimated, as it would be the case for a longer strobe length (>$\SI{100}{\micro\second}$). The final ITS3 sensor will use a layout without strobe-bias crossings and edge-based latching of a hit, ensuring a fake-hit rate independent of acquisition time.

\subsection{Working point validation of the in-pixel front-end}
\label{subsubsec:two_params_scan}

\begin{figure}[!hbt]
    \centering
    \includegraphics[width=0.9\linewidth]{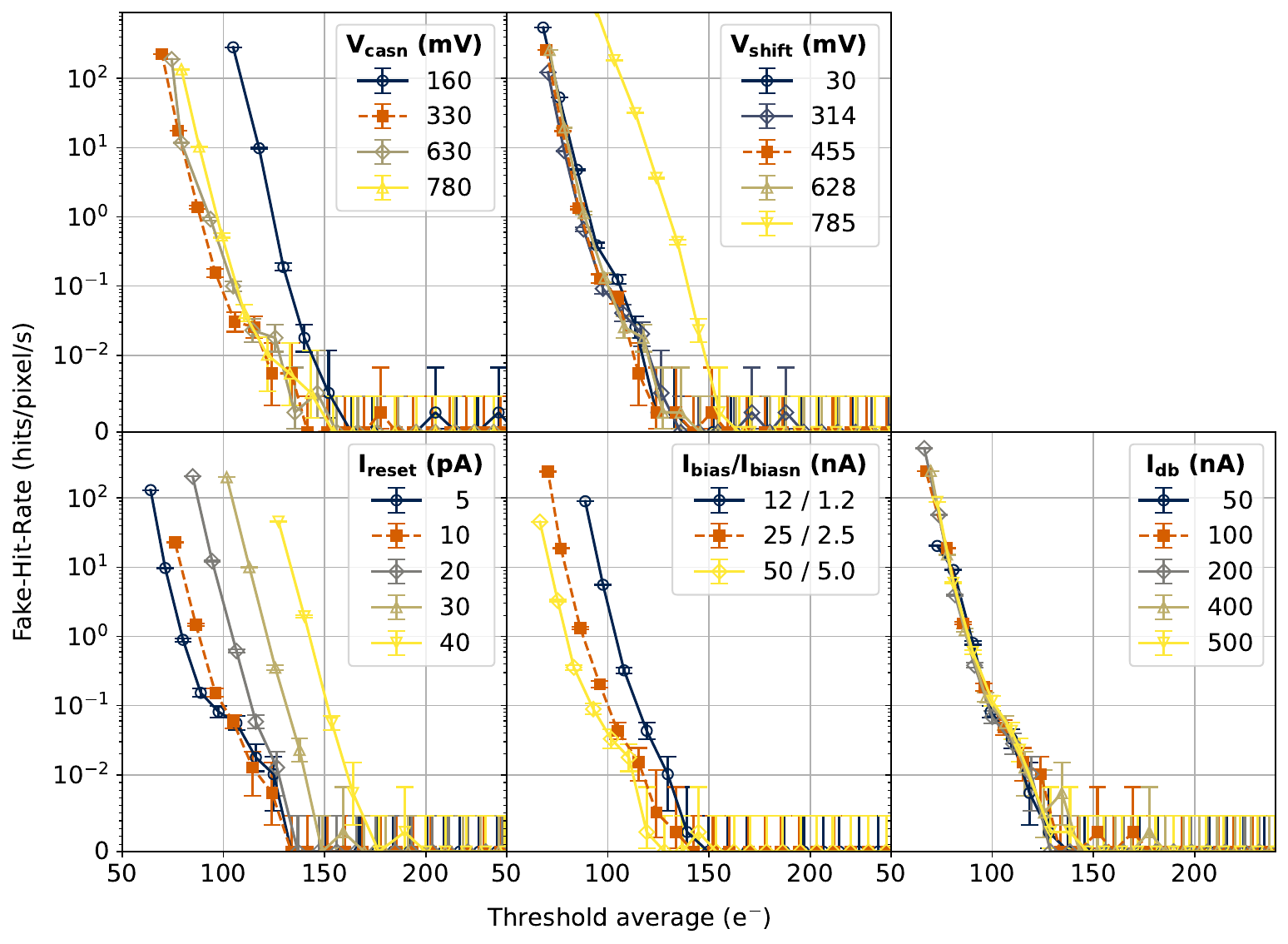}
    \caption{Fake-hit rate as a function of the threshold average per region (set via $\Vb$) for different front-end parameter variations. The plots, arranged from top left to bottom right, show the effect of varying $\Vn$, $\Vs$, $\Ir$, $\Ib$ (together with $\Ibn$ at a ratio 10:1), and $\Id$. The error bars indicate the statistical uncertainty of the fake-hit rate measurement. The nominal operating point extracted from simulations (red dashed lines) was confirmed to yield optimal fake-hit rate performance for a given threshold.}
    \label{fig:two_par_scans}
\end{figure}

The front-end operating point, as a function of bias parameters, was extensively studied to validate the design simulations and optimize the signal-to-noise ratio.
Figure~\ref{fig:two_par_scans} illustrates the relation of the fake-hit rate and the threshold as individual front-end parameters are varied. The parameter $\Vb$ serves as a linear handle to compensate for threshold shifts induced by other front-end settings, allowing consistent threshold tuning without affecting overall front-end operation (see Sec.~\ref{sec:frontend}).

Starting with the parameter $\Ir$, a reduction in the fake-hit rate at a given threshold is observed as $\Ir$ decreases. This behaviour is consistent with an increased front-end gain, attributed to a higher resistance in the reset and feedback branch, and with a lower shot noise from a reduced current injected into the collection electrode. is further supported by the expectation that a lower $\Ir$ reduces shot noise at the collection electrode. Since $\Ir$ also influences the diode biasing, it can be adjusted to compensate for the increased leakage current in sensors exposed to non-ionising radiation. To maintain uniformity in the biasing conditions across different irradiation levels, without compromising the noise performance, a nominal setting of \SI{10}{\pico\ampere} is adopted.

An increase in the front-end bias current, $\Ib$, reduces the fake-hit rate at a given threshold. This is consistent with an increase in the overall front-end voltage gain, and a reduction in the thermal noise of the input transistor M1. Since $\Ib$ is the primary contributor to the power consumption of the in-pixel front-end, a value of \SI{25}{\nano\ampere} is chosen as a trade-off between the ITS3 power budget constraints~\cite{ITS3_TDR} and fake-hit rate performance.

Variations of discriminator current $\Id$, cascode voltage $\Vn$, and level-shifting voltage $\Vs$ around their nominal values (\SI{100}{\nano\ampere}, \SI{330}{\milli\volt}, and \SI{460}{\milli\volt}, respectively) do not lead to significant changes in the observed performance. 
However, at elevated values of $\Vs$, the M0 transistor in the front-end circuit (see Fig.~\ref{fig:frontend}) enters the ohmic region, reducing the front-end gain. Consequently, the fake-hit rate increases for a given threshold. A similar effect is observed for low values of $\Vn$, where the M9 transistor also transitions into the ohmic region, again leading to degraded performance.
While choosing a lower $\Id$ current can reduce dynamic power consumption, it requires a corresponding decrease in $\Vb$ to compensate for the associated threshold reduction (see Sec.~\ref{sec:frontend}). However, this limits the available adjustment range of $\Vb$, particularly for sensors affected by threshold shifts due to ionising radiation, hence, the nominal value was kept at \SI{100}{\nano\ampere}.

Overall, the nominal operating point extracted from simulations was confirmed to yield optimal fake-hit rate performance for a given threshold and is adopted for all subsequent measurements and analyses presented in this work.

\subsection{Time-over-Threshold measurement}
\label{sec:Tot}
\label{sec:pixelcalib}

The Time-over-Threshold (ToT) quantifies the duration for which the signal at the discriminator input exceeds the threshold, thereby encoding the charge information in time.
The ToT of a MOSS front-end signal can be assessed in a specific readout configuration. Here, the strobe signal is set to last much longer than a typical front-end pulse. When charge is collected, the discriminator output activates, leading the hit to be stored in the corresponding in-pixel latch (see Sec.~\ref{sec:moss_matrix_dacs}). This triggers a global signal indicating data presence in the pixel matrix. As soon as this signal is propagated to the test system, a readout command is sent. This initiates the readout of the pixel address(es) by the region periphery and resets the in-pixel latch. As long as the strobe and discriminator output remain active, the pixel-hit latch will immediately reassert and the region readout will read out the same pixel address repeatedly. The number of generated pixel addresses is then proportional to the duration of the front-end signal, thereby providing ToT measurement.

\begin{figure}[!htb]
    \centering
    \includegraphics[width=0.5\linewidth]{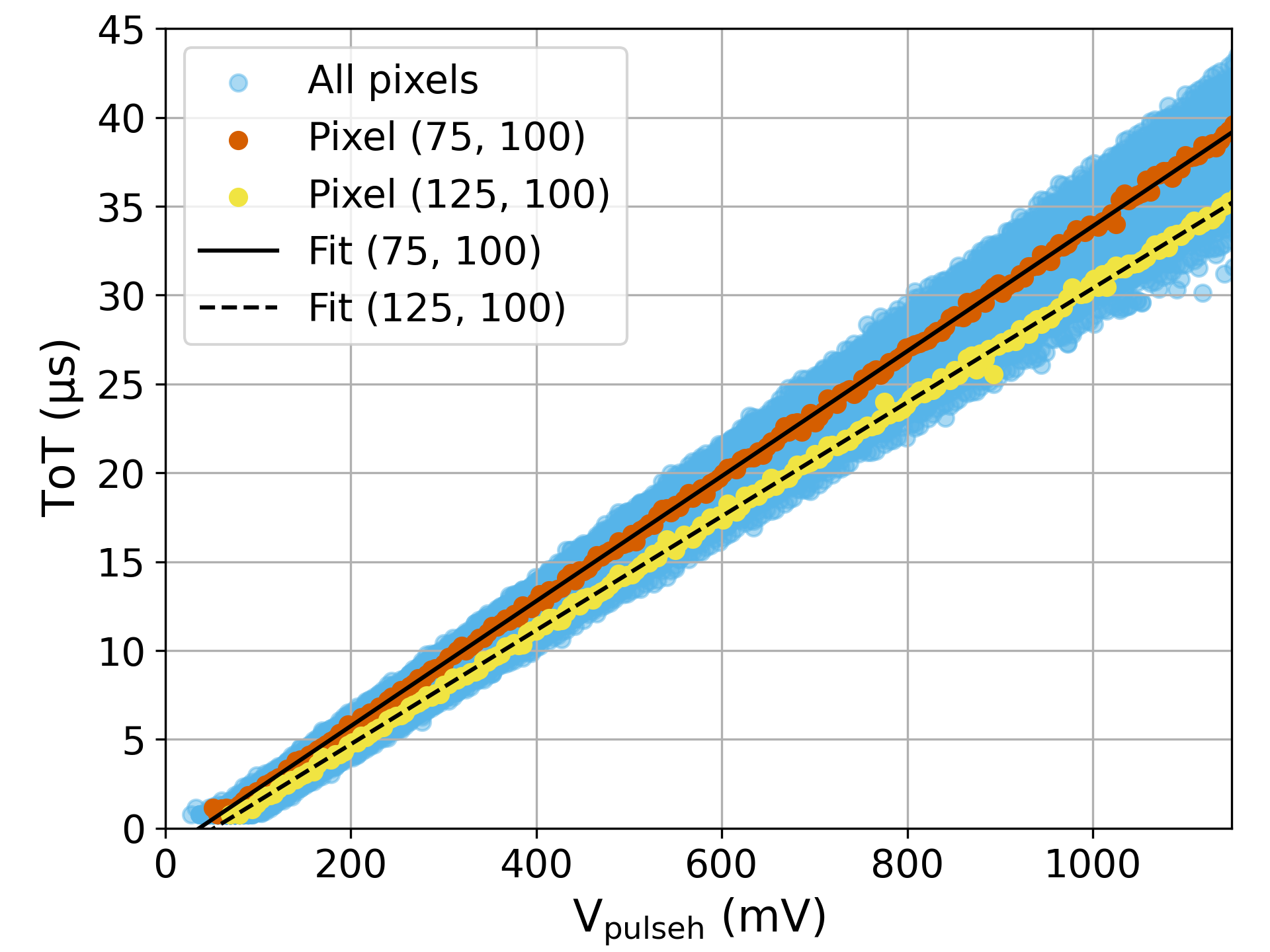}
    \caption{ToT as a function of the pulsing voltage $\Vh$. Two arbitrary pixels with distinct slopes are highlighted to better illustrate pixel-to-pixel variations. Pixels are fitted with a linear function, which is used to calibrate the pixel ToT response.}
    \label{fig:tot_calib}
\end{figure}

The ToT response, similar to the threshold (see Sec.~\ref{sec:threshold_noise_fhr}), exhibits pixel-to-pixel variations due to fabrication-related non-uniformities. Figure~\ref{fig:tot_calib} presents the ToT measured across a number of pixels within a selected region as a function of the pulsing voltage $\Vh$, modulating the injected charge. For each pixel, the ToT response increases linearly with $\Vh$, but the slope and intercept of this relationship vary across pixels.
To ensure a uniform matrix response, each pixel’s ToT response is individually calibrated by fitting a straight line.
This calibration is applied to all the following measurements.

\subsection{Measurements with X-rays}
\label{sec:fe55}

\begin{figure}[!htb]
    \centering
    \includegraphics[width=0.5\linewidth]{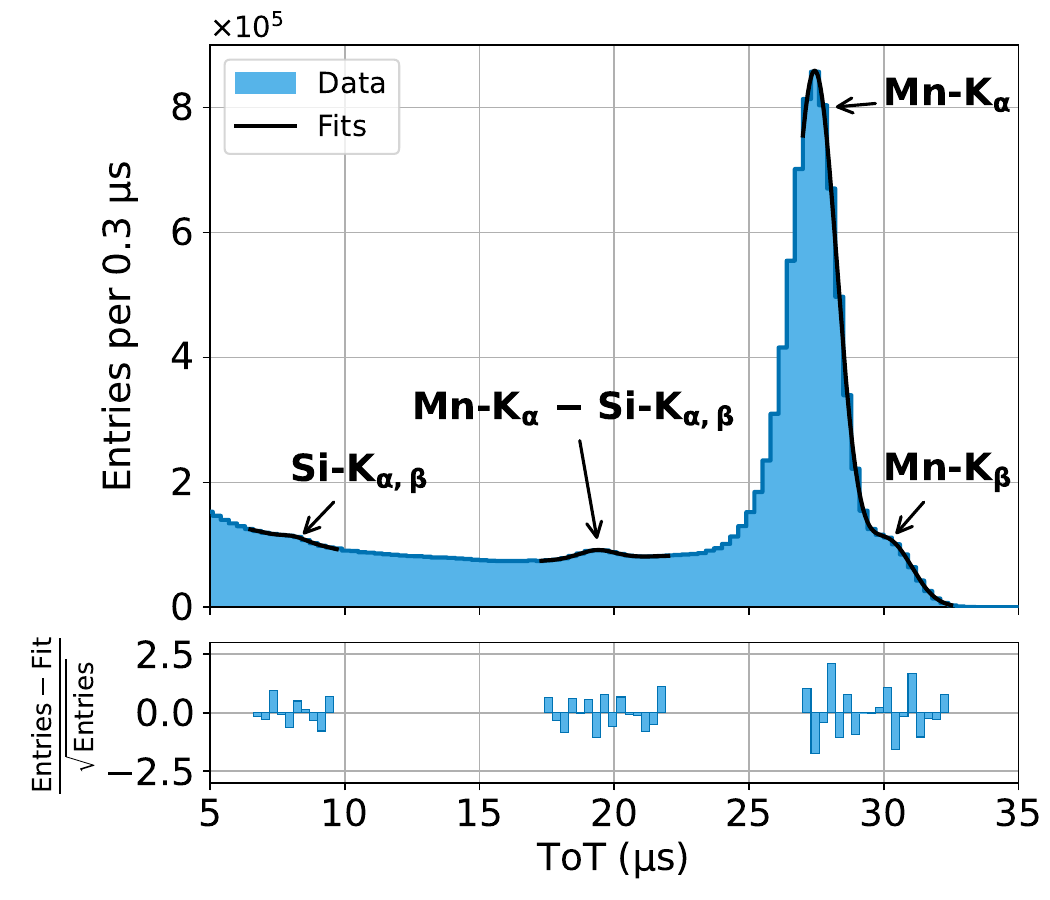}
    \caption{Single-pixel cluster ToT spectrum measured with an $^{55}$Fe source.
    The primary \mnka and \mnkb X-ray emissions are resolved, as well as the secondary \sifl and \siesc{} peaks.
    Fit residuals shown in the bottom plot are within 2.5~standard deviations.}
    \label{fig:fe55}
\end{figure}

Figure~\ref{fig:fe55} shows the sensor response to X-ray emissions from a \textsuperscript{55}Fe, plotted as a ToT spectrum of events where the charge is detected by a single pixel. Besides the primary \SI{5.9}{keV} \mnka and \SI{6.5}{keV} \mnkb emissions from the \textsuperscript{55}Fe decay, spectral features arising from secondary interactions within the silicon sensor are also resolved: namely, the silicon fluorescence line \sifl and the escape peak \siesc. The \mnka and \mnkb peaks are fitted with a sum of two Gaussians. The \sifl and \siesc{} peaks are fitted with a Gaussian added to a linear background. Based on the fit to the dominant \mnka emission line, the energy resolution is determined to be $\textrm{FWHM}/\textrm{Mean}=\SI{7.3(0.2)}{\percent}$, consistent with values reported for the previous prototype~\cite{dpts_paper}.

\begin{figure}[!htb]
    \centering
    \includegraphics[width=0.5\linewidth,trim={0 0.5cm 0 0.5cm},clip]{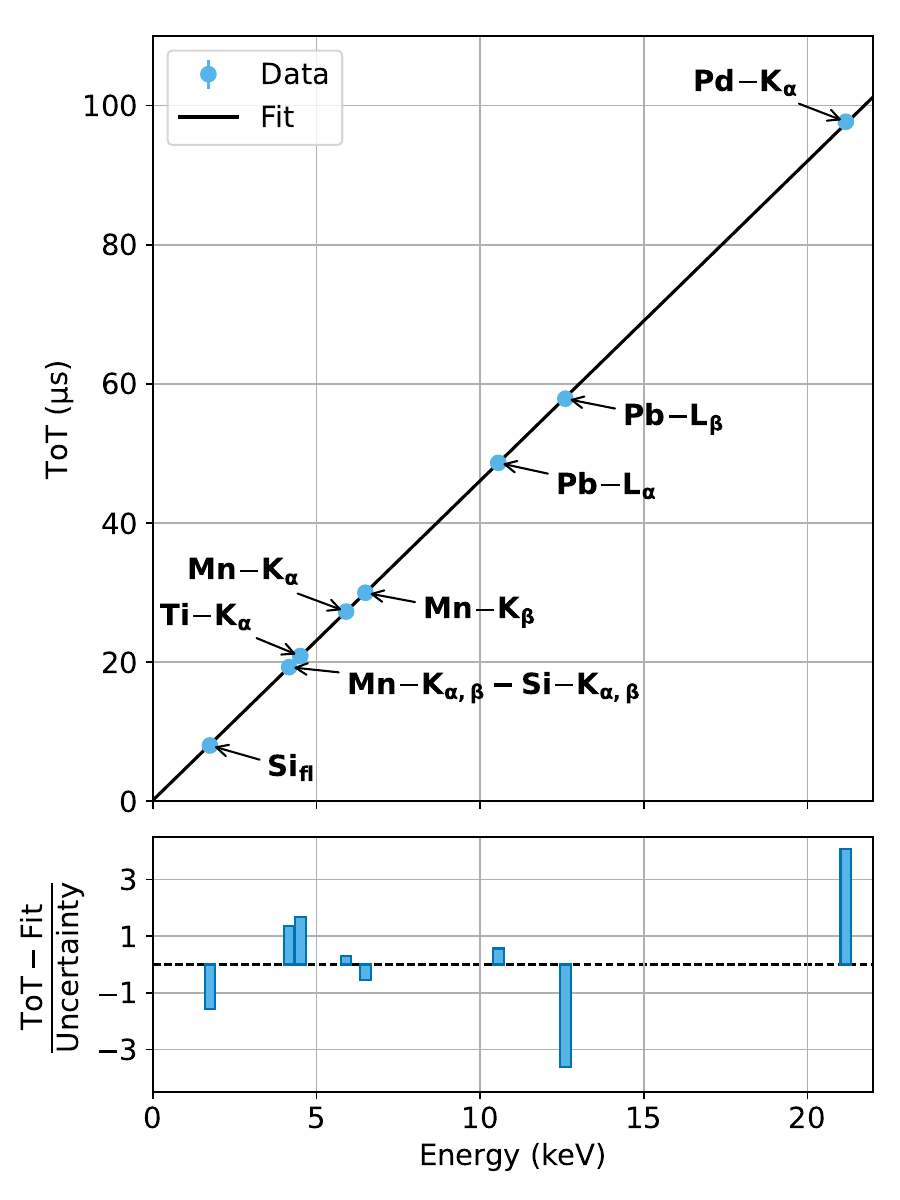}
    \caption{Linearity of the in-pixel front-end response to collected charge. The observed emission line positions are plotted against corresponding literature values, with a linear fit to the data shown in black.}
    \label{fig:ecal}
\end{figure}

X-ray fluorescence emissions from titanium, lead, and palladium were also measured and fitted. The measurement took place at the OptImaTo laboratory~\cite{Trapani_2024}, located at Elettra Sincrotrone Trieste~\cite{elettra}. The experimental conditions were equivalent to those described in Ref.~\cite{dpts_2}.
By comparing the fitted peak positions with literature energy values, the ToT response was confirmed to be linear across the energy range from \SI{1.7}{keV} to \SI{21.2}{keV}, as shown in Fig.~\ref{fig:ecal}.

\subsection{Injection capacitance calibration}
\label{sec:pulsing_calib}

Due to fabrication process variations, the injection capacitance (denoted as $\Cinj$ in Fig.~\ref{fig:frontend}) must be calibrated to enable accurate comparison of threshold values across different pixels and regions. Given the confirmed linearity of the ToT response with both injected charge and deposited energy (see Figs.~\ref{fig:tot_calib} and~\ref{fig:ecal}, respectively), the injection capacitance can be determined by measuring the position of the \mnka peak (see Fig.~\ref{fig:fe55}) and calculating the corresponding pulsing voltage, $\Vh$ (see Fig.~\ref{fig:tot_calib}). The injection capacitance is then obtained as $\Cinj = Q_{\mathrm{Mn\textnormal{-}K}_{\alpha}}/\Vh$, where $Q_{\mathrm{Mn\textnormal{-}K}_{\alpha}}$ is the charge deposited by photoelectric absorption of a \mnka emission.

\begin{figure}[!hbt]
    \centering
    \begin{minipage}{.49\textwidth}
    \includegraphics[width=\linewidth]{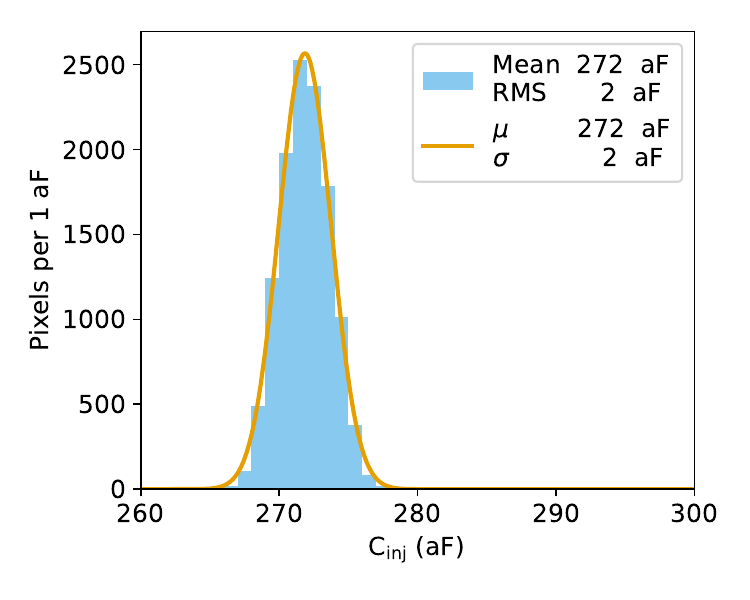}
    \caption{Distribution of injection capacitances measured for a subset of pixels in a non-irradiated region. The orange line represents a Gaussian fit. The measured spread is approximately \SI{1}{\percent} of the mean value. \label{fig:cinj_pixels}}
\end{minipage}%
\hfill
\begin{minipage}{.49\textwidth}
    \centering
    \includegraphics[width=\linewidth]{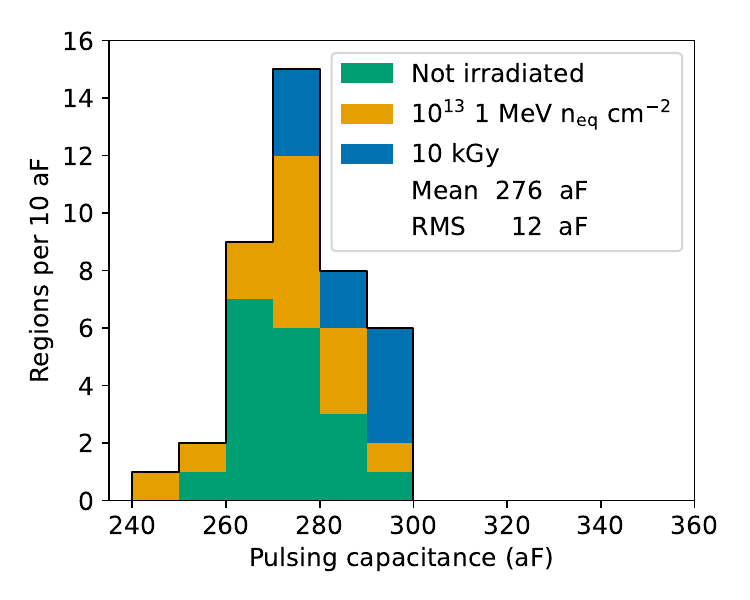}
    \caption{Distribution of the average injection capacitance values for non-irradiated, \SI{10}{\kilo\gray}, and \SI{1e13}{\niel} irradiated regions (stacked). The spread of the distributions reflects region-to-region variations, while no impact of the irradiation is visible.}
    \label{fig:pulsingcapacitance}
\end{minipage}
\end{figure}

Figures~\ref{fig:cinj_pixels} and~\ref{fig:pulsingcapacitance} show the distributions of measured injection capacitance values for a subset of pixels \footnote{To avoid artefacts from non-idealities in the injection circuit~\cite{pulsingcircuitry}, only the central quarter of the region is shown, excluding also affected columns.} in a non-irradiated region and the average values across regions subjected to different irradiation levels, respectively. With an RMS of approximately \SI{1}{\percent} of the mean value, the pixel-to-pixel capacitance spread can be assumed to contribute only marginally to the measured threshold spread (see Sec.~\ref{sec:threshold_noise_fhr}). However, the average capacitance dispersion across different regions cannot be attributed to statistical pixel-to-pixel variation alone and instead indicates systematic region-to-region differences.

To mitigate this spread, the injection capacitance values for all regions presented in this work are calibrated using the described procedure.
The average capacitance of \SI{272(2)}{\atto\farad}, measured across different regions and irradiation levels, is reasonably close to the design value\footnote{The design value of the injection capacitance and the corresponding uncertainty are determined from the typical and extreme parasitic extraction corners.} of \SI{258(22)}{\atto\farad}. The quoted uncertainty corresponds to the statistical uncertainty on the mean across regions, while the design uncertainty reflects variations from parasitic extraction corners. As expected, no impact of irradiation is observed.

\subsection{Detection efficiency and fake-hit rate}
\label{sec:detection_efficiency}

In-beam measurements were conducted to evaluate detection efficiency at different thresholds, using the setup described in Sec.~\ref{sec:testbeam_setup}. The uncertainty of the threshold is calculated by accounting for the statistical uncertainty of the threshold measurement in $\Vh$ DACs (see Sec.~\ref{sec:thr_noi_fhr}) and the statistical uncertainty from the conversion into electrons based on the calibration of the pulsing capacitance (see Sec.~\ref{sec:pixelcalib}). The uncertainty of the detection efficiency and fake-hit rate is evaluated using a Clopper–Pearson interval with a 66.3\% confidence level.

The target performance for ITS3 sensors is a detection efficiency higher than \SI{99}{\percent}, with a fake-hit rate lower than 0.1~hits/pixel/s~\cite{ITS3_TDR}. This performance must be maintained after the radiation doses expected during ITS3 operation, which, including a safety factor, are\footref{fn:radhard} of \SI{4}{\kilo\gray} Total Ionising Dose (TID) and \SI{4e12}{\niel} Non-Ionising Energy Loss (NIEL).

\begin{figure}[!htb]
    \centering
    \includegraphics[width=1\linewidth]{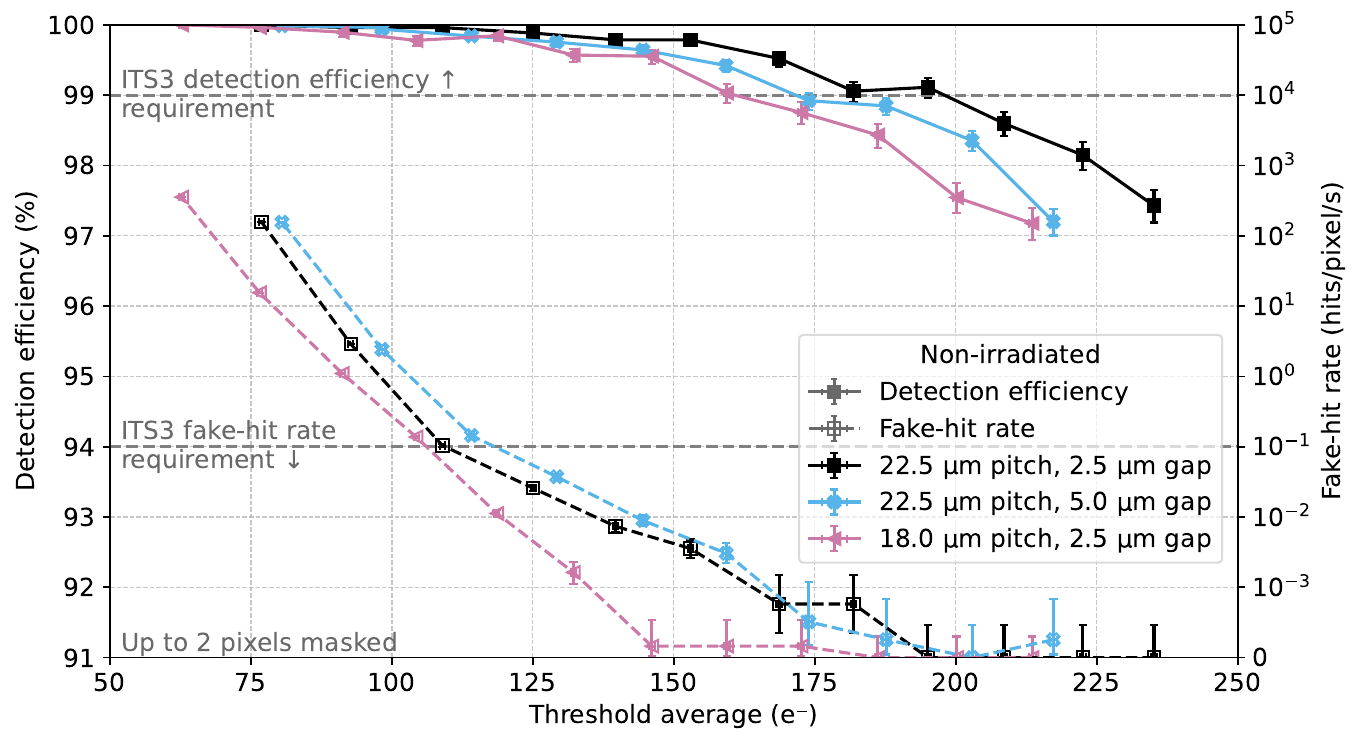}
    \caption{Detection efficiency (filled symbols, solid lines) and fake-hit rate (open symbols, dashed lines) as a function of the average threshold, comparing a non-irradiated \SI{22.5}{\micro\meter} pitch, \SI{18.0}{\micro\meter} pitch, and \SI{22.5}{\micro\meter} pitch with an increased gap size from the \SI{2.5}{\micro\meter} to \SI{5.0}{\micro\meter}.}
\label{fig:detection_efficiency_top_bottom_split_comparison}
\end{figure}

Figure~\ref{fig:detection_efficiency_top_bottom_split_comparison} compares the detection efficiency and the fake-hit rate as a function of the average sensor threshold for two pixel pitches, \SI{22.5}{\micro\meter} and \SI{18.0}{\micro\meter}, and shows the effect of increasing the size of the deep implant gap (see Sec.~\ref{sec:moss_matrix_dacs}) from \SI{2.5}{\micro\meter} to \SI{5.0}{\micro\meter} for the \SI{22.5}{\micro\meter} pitch.
A larger pixel pitch is observed to increase detection efficiency. This is consistent with the fact that larger pixels have proportionally less border area than smaller ones; hits in the border region are statistically more likely to fall below the threshold due to charge sharing between multiple pixels and energy straggling. This effect was previously observed in small-scale prototypes~\cite{apts_paper}.
The pixel matrix with \SI{22.5}{\micro\meter} pixel pitch shows a slightly lower efficiency for a larger gap, in agreement with the expectation that when the gap size is increased, the electric field at the border is decreased, increasing the charge sharing in the corresponding region~\cite{TCAD_simulations}.

The fake-hit rate is higher for the larger pixel pitch. The shot noise from leakage current, which increases with pixel volume, is insufficient to account for this\footnote{Estimated using the thermal model of the fake-hit rate described in Ref.~\cite{dpts_2}.}. This is consistent with the expectation that leakage current is not the dominant contributor in non-irradiated samples. In the smaller-pitch matrix, the higher pixel count increases the capacitive load on the biasing circuit and the number of bias–strobe crossings. This modulates the strobe-induced perturbation (see Sec.~\ref{sec:strobe_and_fhr}), which can explain the different fake-hit rates. Overall, a larger pixel pitch achieves the target performance over a wider threshold range. 

\begin{figure}[!thb]
	\centering
	\begin{subfigure}[t]{0.99\textwidth}
    \includegraphics[width=\textwidth]{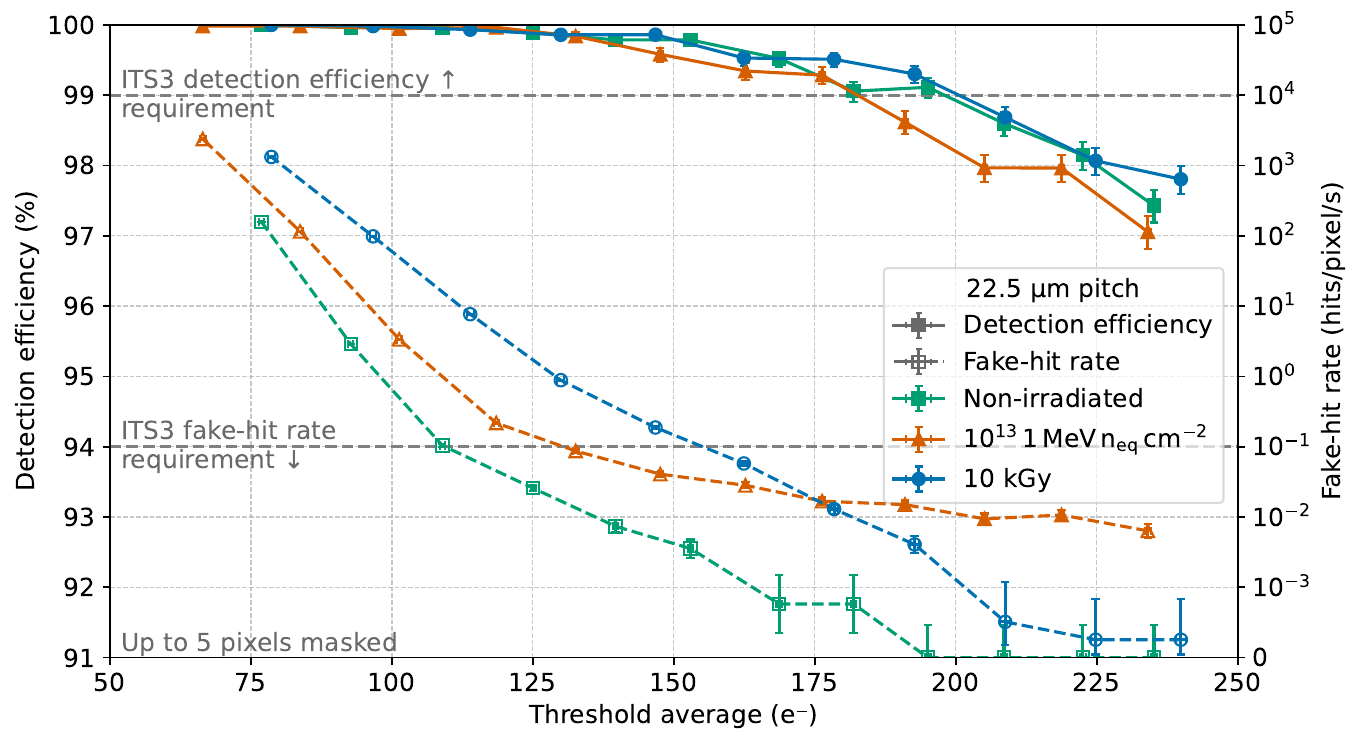}
    \caption{Pixel matrix with pitch \SI{22.5}{\micro\meter} irradiated to different levels.
    }
    \label{fig:detection_efficiency_top}
    \vspace{0.3cm}
	\end{subfigure}
	\begin{subfigure}[t]{0.99\textwidth}
    \includegraphics[width=\textwidth]{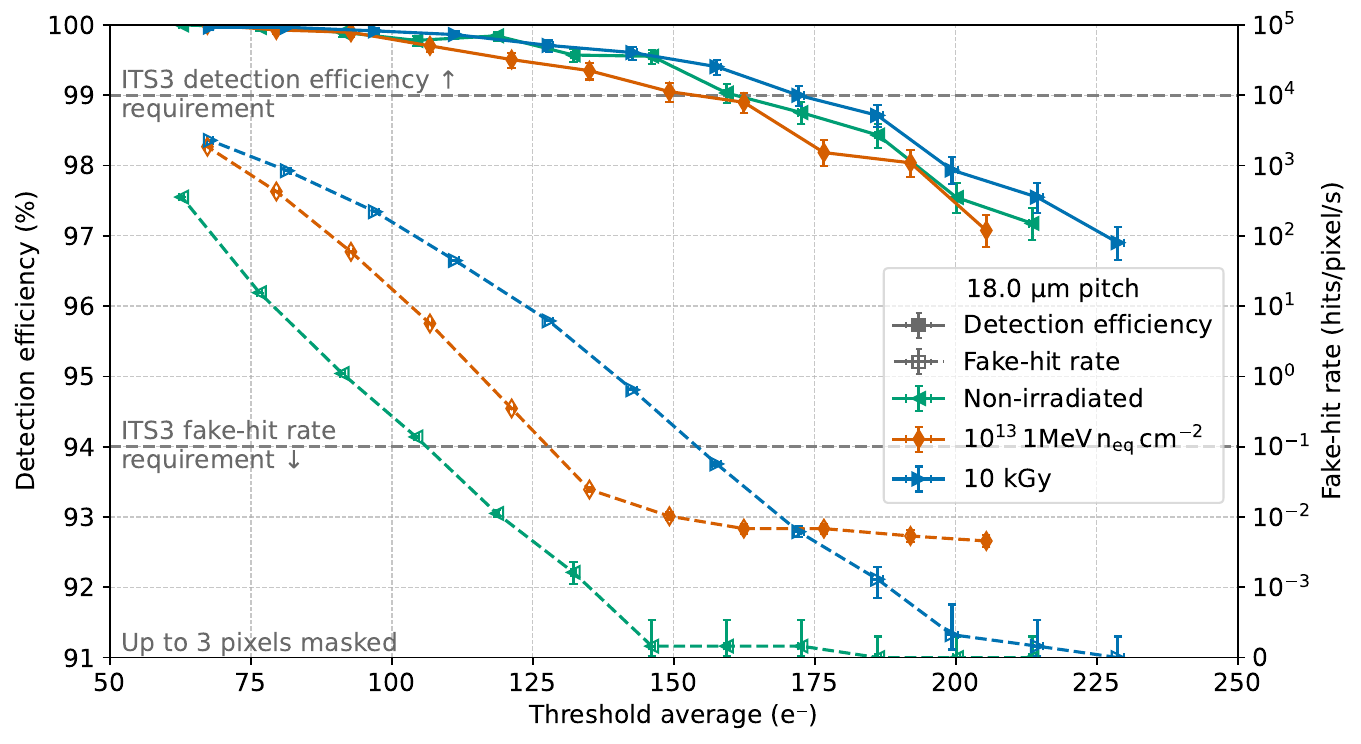}
    \caption{Pixel matrix with pitch \SI{18.0}{\micro\meter} irradiated to different levels.
    }
    \label{fig:detection_efficiency_bottom}
	\end{subfigure}
    \caption{Detection efficiency (filled symbols, solid lines) and fake-hit rate (open symbols, dashed lines) as a function of average threshold for different irradiation levels: non-irradiated, an ionising radiation dose of $\SI{10}{kGy}$ and a non-ionising radiation fluence of \SI{1e13}{\niel}.}
    \label{fig:efficency_fhr_irrad}
\end{figure}

Figures~\ref{fig:detection_efficiency_top} and \ref{fig:detection_efficiency_bottom} show detection efficiency and fake-hit rate for \SI{22.5}{\micro\meter} and \SI{18.0}{\micro\meter} pitches with \SI{2.5}{\micro\meter} gap, as a function of the threshold, for different irradiation levels: non-irradiated, an ionising radiation (TID) dose\footnote{\label{fn:tid}The TID irradiation was done using the CERN Xray machine. The sensors were tested within a week after receiving the ionising radiation dose and again after six months of annealing at room temperature. Comparable results were obtained. The results after the annealing are shown, as they were obtained under the same conditions as the other results in this work.} of \SI{10}{\kilo\gray}, and a non-ionising radiation (NIEL) fluence\footnote{\label{fn:niel}The NIEL irradiation was done using neutrons from the JSI TRIGA Mark II reactor in Ljubiljana. The sensor irradiated with a NIEL fluence was kept at $\SI{-20}{\celsius}$ between the irradiation and the testbeam in order to avoid annealing, as the process of procuring new samples required a significant time and labour investment.} of \SI{e13}{\niel}. The use of Xray photons and neutrons for TID and NIEL irradiation, respectively, instead of charged hadrons was motivated in order to study the impact of both types of irradiation separately.
After irradiation, the operational margin, defined as the threshold range where ITS3 requirements are met, is reduced as expected. Whereas, for example, for the non-irradiated pixel matrix with pitch \SI{22.5}{\micro\meter} shown in Fig.~\ref{fig:detection_efficiency_top_bottom_split_comparison}, the operational margin is between \SI{110}{\ele} and \SI{200}{\ele} for a total range of about \SI{90}{\ele}, it reduces to less than \SI{50}{\ele} at the radiation levels used in this study (see Fig.~\ref{fig:detection_efficiency_top}). For the pixel matrix with a \SI{18.0}{\micro\meter} pitch, the observed operational margin of less than \SI{20}{\ele} (see Fig.~\ref{fig:detection_efficiency_bottom}) can practically be considered as disappeared.

The sample irradiated with non-ionising radiation shows an increased fake-hit rate, expected due to higher sensor leakage from irradiation damage and the associated rise in shot noise. Furthermore, the fake-hit rate plateau at higher thresholds is consistent with the residual radioactivity of the sensors after irradiation. The TID-irradiated sample also shows a higher fake-hit rate, as expected from ionising radiation impacting transistor performance and thus the front-end signal-to-noise ratio. 
Both pixel pitches exhibit comparable trends across different irradiation levels.

\subsection{Spatial resolution and average cluster size}
\label{sec:spatial_resolution}

A particle hit on the sensor may cause one or more pixels to register a signal above threshold. Adjacent pixels registering a hit form a cluster, whose size corresponds to the number of pixels above threshold.
To calculate the spatial resolution, the RMS of the residual (the distances between the track intercept on the device under test and the cluster centre of mass\footnote{with equal weights on each pixel since no analogue information is available}) is computed in both column and row directions. The spatial resolution for the two directions is then obtained by quadratically subtracting the estimated telescope tracking resolution at the DUT position, which is about \SI{2}{\um}. The spatial resolution referred to in the remainder of this paper is the average of the resolutions along the column and row directions. The uncertainty of the spatial resolution is derived from the statistical uncertainty of the RMS of the residual. The error on the cluster size is a statistical error on the mean.

\begin{figure}[!htb]
    \centering
    \includegraphics[width=1\linewidth]{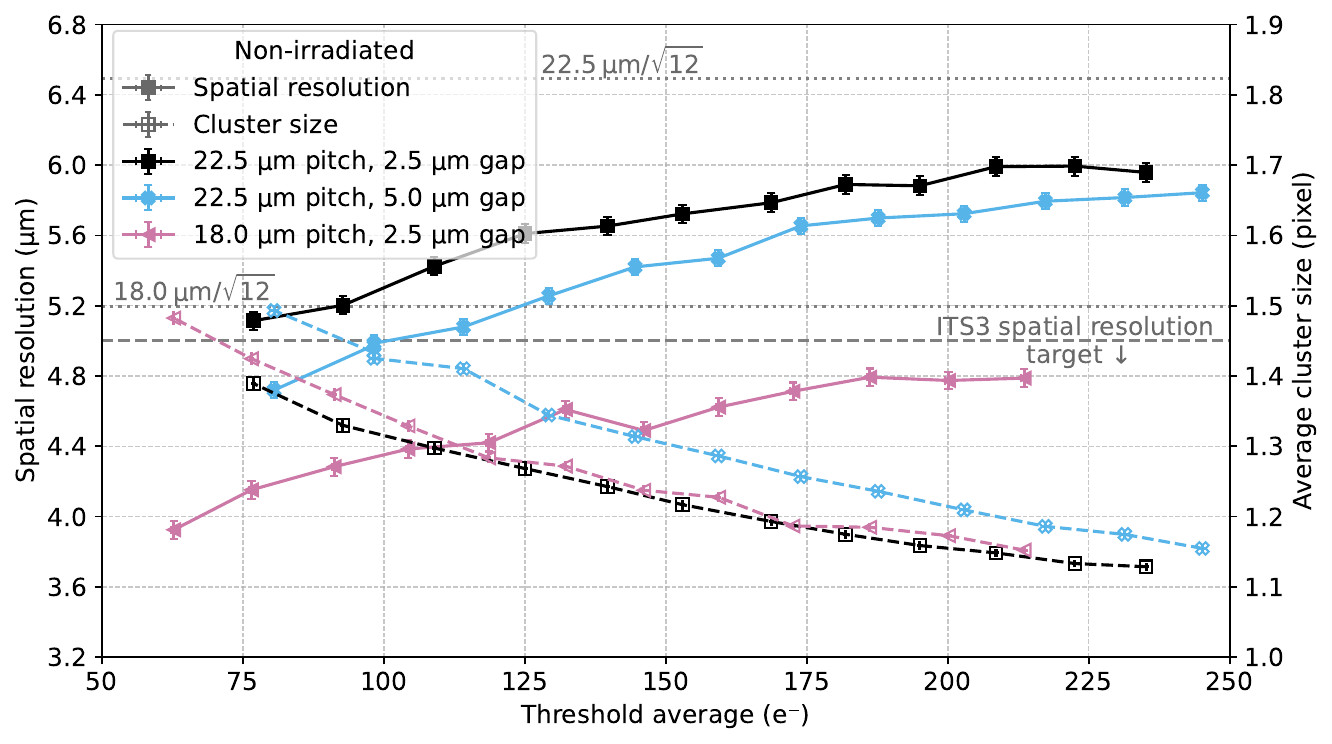}
    \caption{Spatial resolution (filled symbols, solid lines) and average cluster size (open symbols, dashed lines) as a function of the average threshold, comparing a non-irradiated \SI{22.5}{\micro\meter} pitch, \SI{18.0}{\micro\meter} pitch and \SI{22.5}{\micro\meter} pitch with an increased gap size from the \SI{2.5}{\micro\meter} to \SI{5}{\micro\meter}.}
    \label{fig:spatial_resolution_top_bottom_split_comparison}
\end{figure}

Figure~\ref{fig:spatial_resolution_top_bottom_split_comparison} compares spatial resolution and average cluster size for \SI{22.5}{\micro\meter} and \SI{18.0}{\micro\meter} pixel pitches and shows the effect of increasing the gap size for the \SI{22.5}{\micro\meter} pitch.
The ITS3 spatial-resolution target value of $\SI{5}{\micro\meter}$~\cite{ITS3_TDR} is shown as a dashed line. The dotted lines represent the ``hit/no-hit resolution'', expected if the deposited charge were collected by a single pixel. The measured spatial resolution is consistently better than this limit, showing a decreasing trend toward lower thresholds, where larger cluster sizes are observed.
Hits near pixel borders, where charge is shared between multiple pixels, often fall below threshold and are not registered, reducing the average cluster size at higher thresholds.
Across all thresholds, the \SI{18.0}{\micro\meter} pitch provides better spatial resolution due to the finer size. At a threshold of \SI{160}{\ele} (the lowest threshold within the operational range after irradiation, see Fig.~\ref{fig:efficency_fhr_irrad}), the \SI{18.0}{\micro\meter} pitch achieves a spatial resolution of about \SI{4.5}{\micro\meter}, satisfying the ITS3 requirements. In contrast, the \SI{22.5}{\micro\meter} pitch reaches only about \SI{5.7}{\micro\meter}.

Comparing the pixels with a \SI{2.5}{\micro\meter} and a \SI{5}{\micro\meter} gap sizes shows that a larger gap increases the average cluster size, especially at low thresholds. At higher thresholds, the cluster size tends to converge to that of the \SI{2.5}{\micro\meter} gap, as the threshold becomes too high to detect shared signals and border effects contribute less to the average cluster size.
The increased cluster size leads to an improvement of spatial resolution by \SIrange{0.1}{0.4}{\micro\meter} at thresholds below \SI{160}{\ele}, resulting in about \SI{5.4}{\micro\meter} at a threshold of \SI{160}{\ele}. While a \SI{5}{\micro\meter} gap shows potential to enhance spatial resolution, the $\SI{22.5}{\micro\meter}$ pixel pitch remains above the ITS3 target value.

\begin{figure}[!thb]
	\centering
	\begin{subfigure}[t]{0.99\textwidth}
    \includegraphics[width=\textwidth]{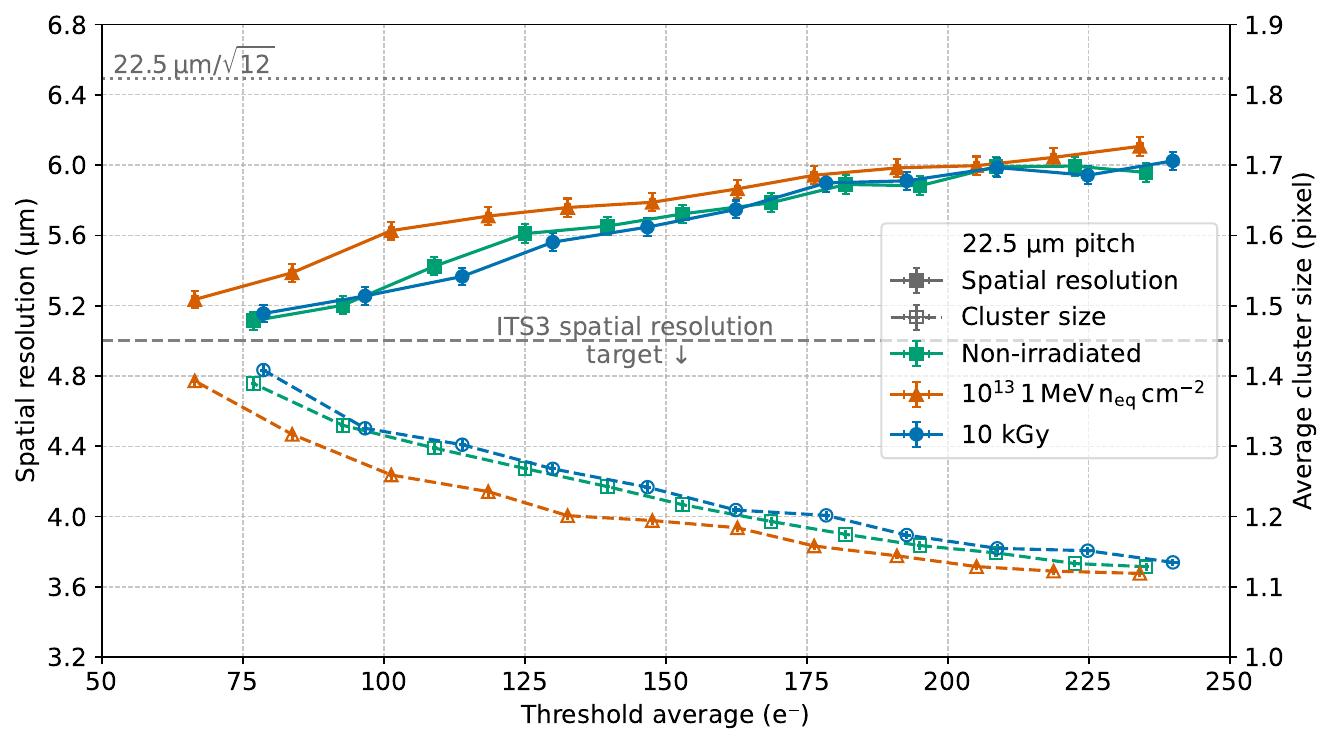}
    \caption{$\SI{22.5}{\micro\meter}$ pixel pitch.
    }
    \label{fig:spatial_resolution_top}
    \vspace{0.3cm}
	\end{subfigure}
	\begin{subfigure}[t]{0.99\textwidth}
    \includegraphics[width=\textwidth]{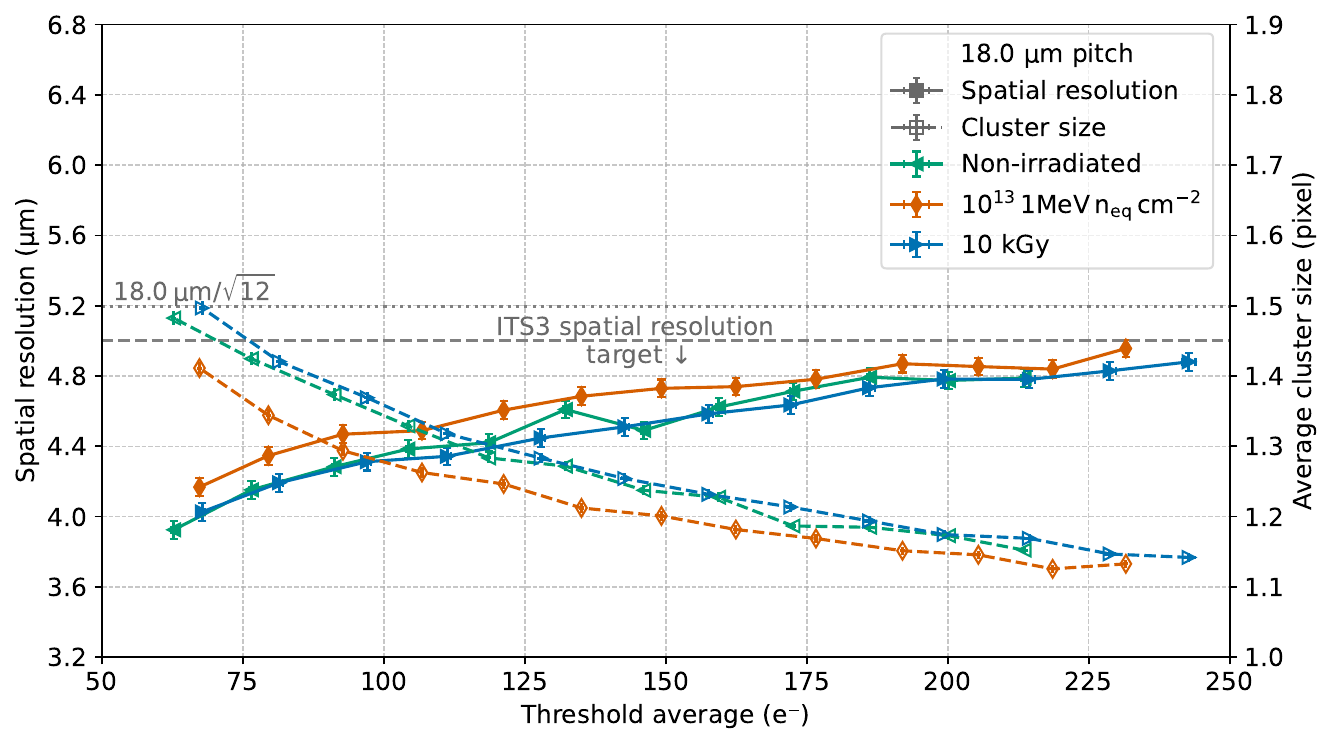}
    \caption{$\SI{18.0}{\micro\meter}$ pixel pitch.
    }
    \label{fig:spatial_resolution_bottom}
	\end{subfigure}
    \caption{Spatial resolution (filled symbols, solid lines) and average cluster size (open symbols, dashed lines) as a function of the average threshold for different irradiation levels: non-irradiated, an ionising radiation dose of $\SI{10}{kGy}$ and a non-ionising radiation fluence of \SI{1e13}{\niel}.}
\end{figure}

Figures~\ref{fig:spatial_resolution_top} and \ref{fig:spatial_resolution_bottom} show the spatial resolution and average cluster size for $\SI{22.5}{\micro\meter}$ and $\SI{18.0}{\micro\meter}$ pitches with a $\SI{2.5}{\micro\meter}$ gap, as functions of the average threshold for different irradiation levels: non-irradiated, an ionising radiation dose\footref{fn:tid} of $\SI{10}{kGy}$, and a non-ionising radiation fluence\footref{fn:niel} of \SI{1e13}{\niel}.
After non-ionising irradiation, the average cluster size decreases at low thresholds, causing a corresponding degradation in spatial resolution of up to $\SI{0.3}{\micro\meter}$. 
At higher thresholds, no significant difference is observed between non-irradiated and NIEL-irradiated sensors, as signals from charge sharing at pixel borders already fall below threshold even without irradiation.
Since ionising radiation affects mostly the in-pixel front-end, shifting the threshold and noise without impacting charge collection, cluster size or spatial resolution shows no significant effect caused by the delivered ionising dose.
\section{Stitched design validation}
\label{sec:series_testing}

To validate the stitched sensor design and evaluate production yield, a test campaign was carried out on 82 non-irradiated MOSS sensors from 14 wafers. This section summarises the results in terms of the yield of different sensor components and the overall yield of wafer-scale sensors.
The tests were performed by powering, controlling, and reading out the MOSS sensor via the long edge, one half-unit at a time (see~Sec.~\ref{subsec:architecture}). Half-units that passed these tests were subsequently retested via the LEC (short edge) to verify communication across the stitching boundary.
The test procedure begins by powering the half-units, followed by verifying the functionality of digital and analogue periphery, and concludes with characterizing the pixel matrix. Testing was conducted in a laboratory with centrally controlled air temperature. The sensor temperature was continuously monitored but not actively controlled. The \SI{4}{\celsius} maximum temperature difference between sensors recorded across tests is not considered to have a significant impact on the interpretation of the following results (see Sec.~\ref{sec:threshold_noise_fhr}).

\subsection{Powering}
\label{sec:powering}

\begin{figure}[!htb]
    \centering
    \includegraphics[width=\linewidth]{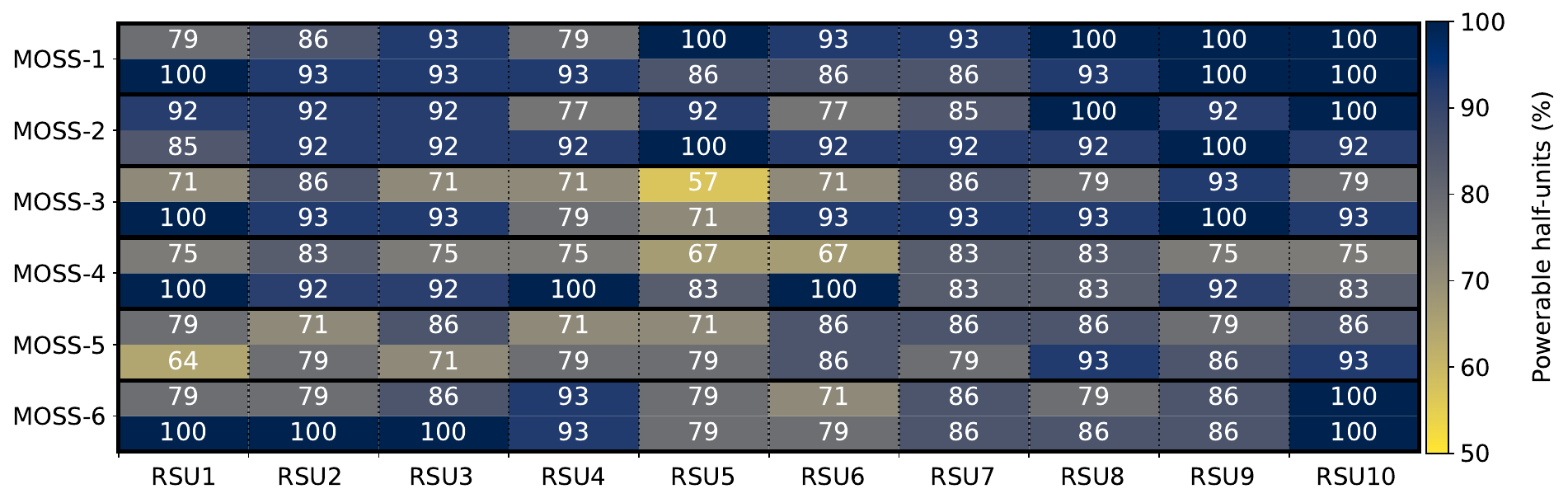}
    \caption{Fraction of half-units that can be powered, as a function of the location on the six MOSS sensors on each wafer. A radial gradient is observed, with reduced yield concentrated near the wafer centre. A detailed investigation correlated the failures with features of the new metal stack~\cite{metal_stack_paper}.}
    \label{fig:powerable_HUs_percent}
\end{figure}

During the powering test, each net of a half-unit was activated sequentially, followed by the application of a clock signal, a reset procedure, and the configuration of the nominal operating point.
The spatial distribution of successfully powered half-units is visualized as a wafer map in Fig.~\ref{fig:powerable_HUs_percent}. 
A radial gradient is observed in the powering yield, with reduced functionality concentrated near the wafer centre. 
The incidence of these faults varies significantly across wafers, as will be shown in Sec.~\ref{sec:yield}.

These patterns suggest underlying manufacturing issues.
The MOSS design introduced a novel metal-stack configuration, implemented for the first time during the ER1 production run. This configuration was specifically customized by the foundry in a collaborative effort to meet the requirements of the ITS3 sensor development project. A detailed investigation, reported in Ref.~\cite{metal_stack_paper}, correlated the failures with features of the new metal stack and facilitated the implementation of corrective measures by the foundry. The findings also provided valuable insights for future design iterations aimed at mitigating similar risks.

\begin{figure}[!htb]
    \centering
    \includegraphics[width=0.9\linewidth]{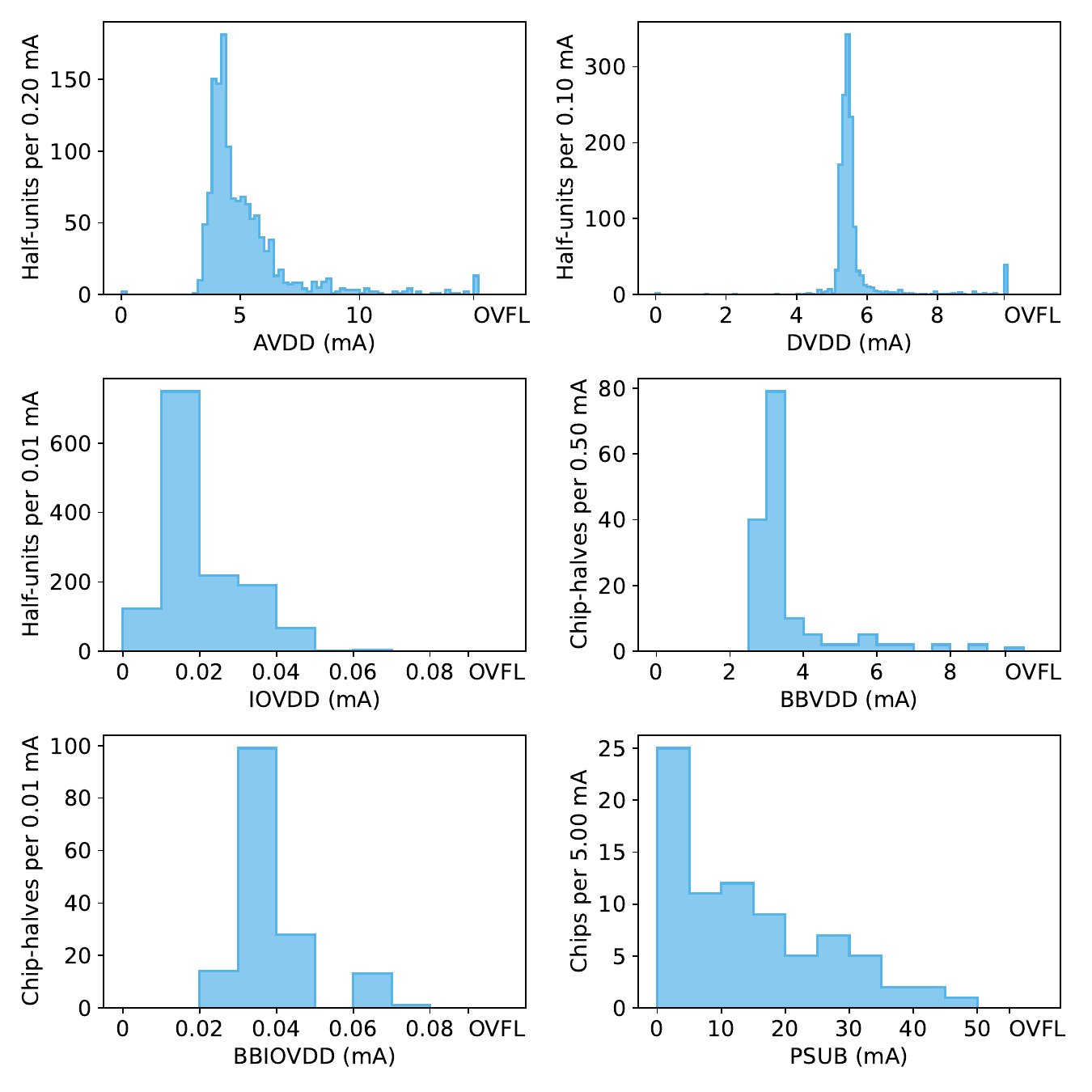}
    \caption{Current distribution for all power nets across all successfully powered sensors. AVDD, DVDD, and IOVDD nets are measured per half-unit (1353 entries). BBVDD and BBIOVDD nets are measured for the top and bottom half of each sensor (145 entries). The PSUB current is measured per MOSS sensor (76 entries). Currents above overflow (`OVFL') are not shown for better visualisation of the individual distributions.}
    \label{fig:endpoint_currents_power_on}
\end{figure}

The current on each power net is measured after all powerable half-units and the LEC are switched on and configured. The distributions of current values are shown in Fig.~\ref{fig:endpoint_currents_power_on}.
These data indicate the typical operating conditions of the sensor and provide a baseline for understanding sensor-to-sensor performance differences (see Sec.~\ref{sec:pix_performance}).
Power nets AVDD, DVDD, and IOVDD are measured per half-unit. Power nets BBVDD and BBIOVDD are measured for the top and bottom half of each sensor. The PSUB current is global and measured per MOSS sensor.
The larger current spread in the analogue net (AVDD) is attributed to differing consumption between the top and bottom half-units (with lower and higher pixel density, respectively), and to varying region front-end implementations at default biasing conditions. Both digital level-shifting nets, IOVDD and BBIOVDD, show the expected low current consumption when no signals are being transferred.
All cases, both within the plotted ranges and in the overflow bin, remain fully operational with the increased current attributable to the aforementioned metal-stack issue.
Similarly, cases on the opposite side of the AVDD and DVDD spectra, with notably lower current, occur when current from one net is sunk by the other. The large spread in substrate current (PSUB) is also attributed to the same issue, and, additionally, to protection diode structures between power nets.

\subsection{Digital periphery tests}
\label{sec:dig_periphery_tests}

Successfully powered half-units are tested for proper functionality in the digital periphery, including slow control communication and register access. The test involves writing and reading back different patterns to all registers\footnote{Each half-unit contains 402 registers. Registers that could place the sensor in an unstable condition, for example by setting the pixel front-end to extremely high power consumption, were only read out and not written to.}.
In total, \SI{0.1}{\percent} of regions exhibited one or more register readback errors. Since these errors affected only a single out of four regions in a half-unit, they are attributed to defects within the sensor periphery rather than slow-control communication.
Additionally, two shift-registers used for masking pixels were tested, with \SI{0.1}{\percent} of regions showing failures. These failures are not considered critical, since the structure was implemented solely to simplify the prototype design and will not be included in the final ITS3-sensor design. The radiation sensitivity of these registers is discussed in Sec.~\ref{sec:see_seu}.

\subsection{Analogue biasing}
\label{sec:analogue_biasing}

Testing of the analogue biasing block begins with tuning the bandgap reference voltages (see Sec.~\ref{sec:moss_matrix_dacs}), followed by measuring the DAC reference voltages and currents. Each DAC (8 per region, 640 per MOSS sensor) is then varied over its full range, and its output is measured via external ADCs connected to dedicated bonding pads on the long edge of the sensor. The purpose of this test is to verify that the in-pixel front-end can be biased within the designed range.

\begin{figure}[!hbt]
     \centering
     \begin{subfigure}[]{0.45\textwidth}
         \includegraphics[width=\textwidth]{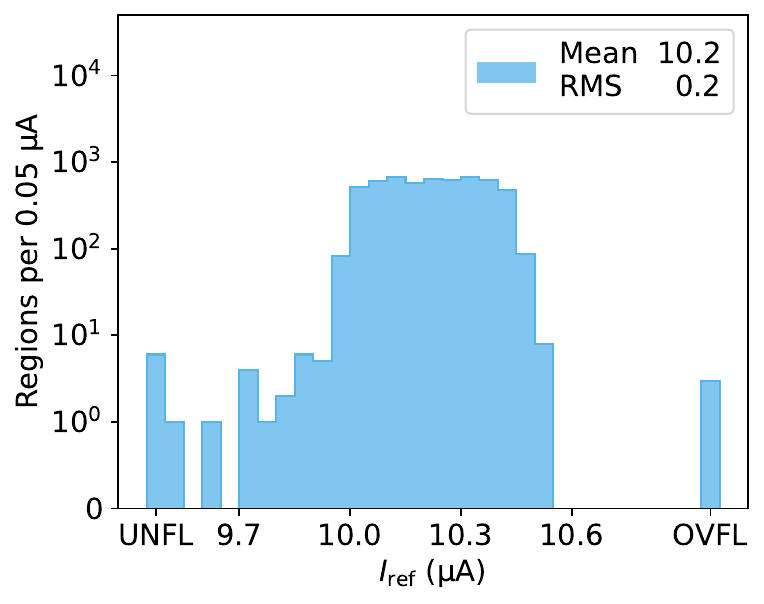}
         \caption{Reference currents.}
    \label{fig:dac-iref}
    \end{subfigure}
    \hspace{0.05\textwidth}
    \begin{subfigure}[]{0.45\textwidth}
         \includegraphics[width=\textwidth]{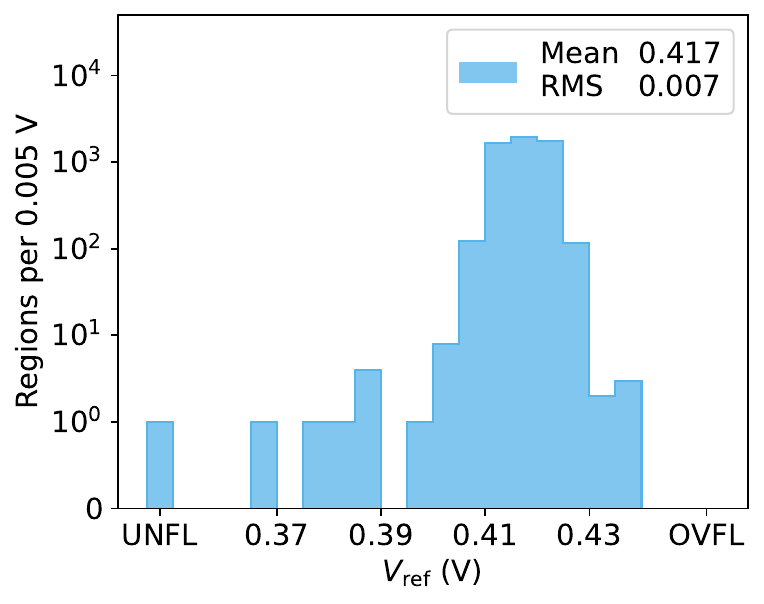}
        \caption{Reference voltages.}
        \label{fig:dac-vref}
    \end{subfigure}
    \caption{Distributions of reference currents and voltages for all regions, with values outside the range placed in underflow and overflow bins. The average values are close to the target values of \SI{10.2}{\micro\ampere} and \SI{0.4}{\volt} for $\Iref$ and $\Vref$, respectively.}
\label{fig:dac-vref-iref}
\end{figure}

The distributions of reference currents and voltages for all regions are shown in Fig.~\ref{fig:dac-vref-iref}. The average values are close to the target values of \SI{10.2}{\micro\ampere} and \SI{0.4}{\volt} for $\Iref$ and $\Vref$, respectively. The spread, depending also on the limited precision of the bandgap tuning, is small enough (FWHM less than \SI{5}{\percent}) to provide reliable biasing references across regions and sensors. The few outliers are not expected to affect sensor operation, as their impact can be compensated by adjusting the corresponding DAC settings.

\begin{figure}[!hbt]
     \centering
     \begin{subfigure}[]{0.49\textwidth}
         \includegraphics[width=\textwidth]{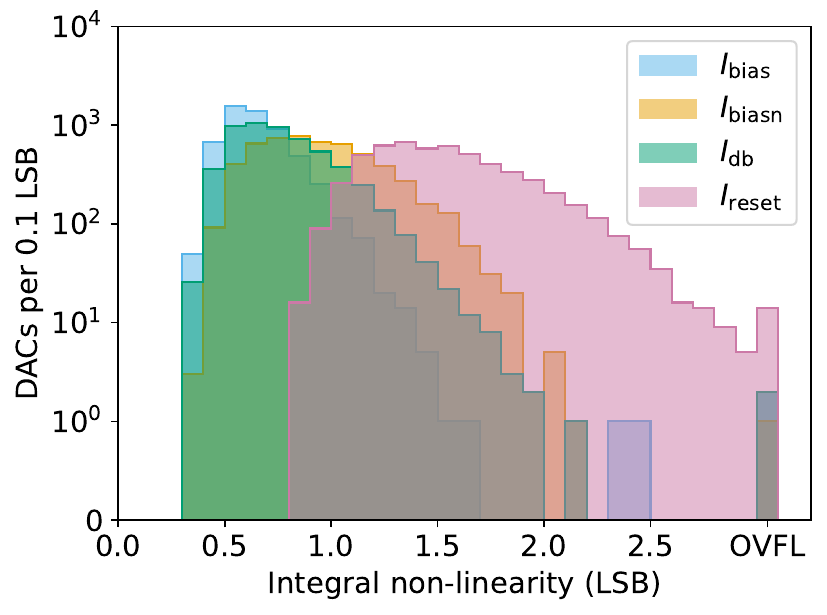}
         \caption{Current DACs.}
    \label{fig:dac-inl_i}
     \end{subfigure}
    \begin{subfigure}[]{0.49\textwidth}
         \includegraphics[width=\textwidth]{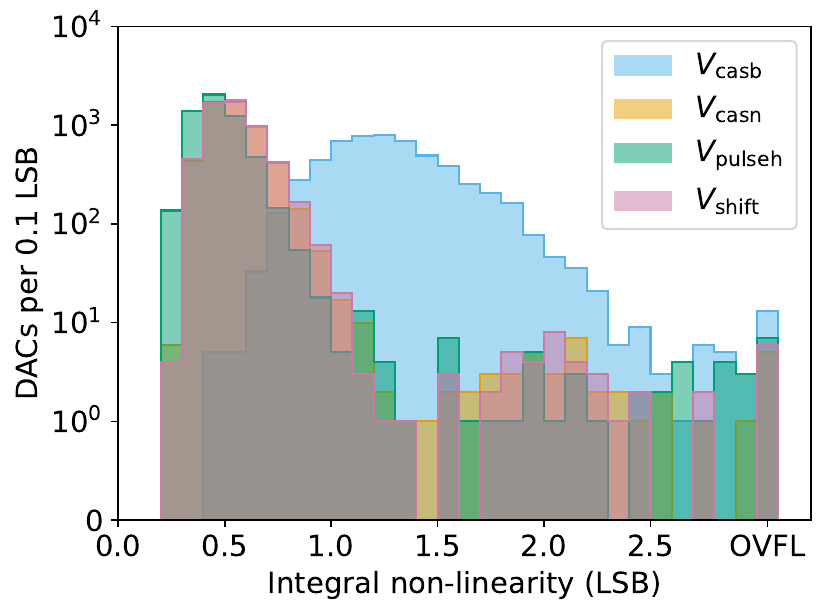}
        \caption{Voltage DACs.}
       \label{fig:dac-inl_v}
     \end{subfigure}
    \caption{Distributions of DAC integral non-linearity for all tested regions. DACs with integral non-linearity exceeding 3~DAC counts are considered non-compliant.}
\label{fig:dac-inl}
\end{figure}

An analysis was performed for each DAC to assess linearity and operational range. As an example, Fig.~\ref{fig:dac-inl} shows the distribution of integral non-linearity for all measured current and voltage DACs. In general, linearity stays within 3~DAC counts for the 8-bit DACs (see Sec.~\ref{subsec:architecture}). The same result was observed for the differential non-linearity. DACs with integral non-linearity above 3~counts were considered non-compliant. $\Ir$ and $\Vb$ DACs show larger non-linearity than the others. For $\Ir$, this is due to the difficulty of supplying and measuring very small currents of the order of a few picoamperes. For $\Vb$, changing its value in a critical range can strongly affect the operating point, altering the power consumption of the pixel matrix and causing supply voltage drops that affect the voltage measurement itself.

A region passes the analogue biasing test if all its DACs perform within predefined ranges, ensuring that reliable biasing conditions of the pixel matrix can be established within the operational range (see Sec.~\ref{subsubsec:two_params_scan}). In total, \SI{0.65}{\percent} of regions did not meet this criterion.


\subsection{Pixel matrix readout}
\label{sec:matrix_readout}

Pixel matrix readout testing verifies the correct propagation of hit pixel addresses from the pixel matrix to the digital circuitry in the sensor periphery and the successful transmission of data packets to the acquisition system.
First, using dedicated testing circuitry, in-pixel latches are asserted (see Sec.~\ref{sec:moss_matrix_dacs}), and pixel addresses are read out row by row. This stage identifies half-units with faulty data-readout interfaces (\SI{0.1}{\percent} of tested half-units) as well as various pixel matrix issues, including faulty pixels, rows, and columns.
The faulty columns and rows arise from the simplistic matrix-steering and readout architecture adopted in the MOSS sensor, given that its design did not target to achieve readout performance or resilience to faults. Specifically, a pixel whose latch cannot be de-asserted blocks the readout at its address, and if such a pixel also cannot be masked, it forces masking of an entire column, row, or of a full region. Overall, 449 out of 5544~regions (\SI{8.1}{\percent}) had to be fully masked and excluded from readout. Since these failures were anticipated at the design stage, and the final ITS3 sensor will employ a different readout architecture, they are not considered critical.

\begin{figure}[!htb]
    \centering
    \includegraphics[width=0.4\linewidth]{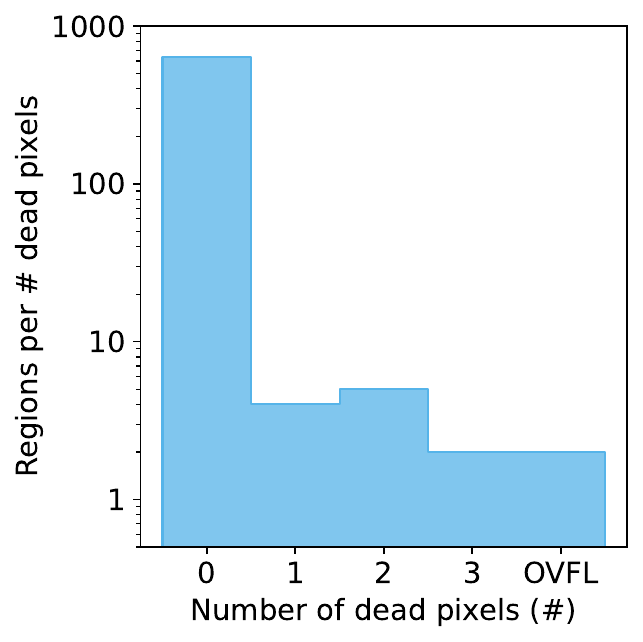}
    \caption{Distribution of dead pixels per region selecting only the top region~0. The other regions exhibit nearly identical behaviour. In most regions, there are no dead pixels.}
    \label{fig:dead_pix_analogue}
\end{figure}

The operation of the pixels is tested in two phases. First, the in-pixel hit latches are set via digital configuration, overriding the analogue front-end and validating the functionality of the digital pixel section and of the matrix readout. Then, testing with the injection of a test charge at the input of the front-end is executed (see Sec.~\ref{sec:frontend}).
Pixels are first classified into four categories with the direct digital test: dead pixels (never fire or always fire, requiring masking), noisy pixels (fire when not expected), inefficient pixels (do not fire reliably when expected), and good pixels. Good pixels are further tested with the charge injection to verify the functionality of the analogue in-pixel front-end. Based on the response to this test, these pixels are reclassified accordingly. As illustrated in Fig.~\ref{fig:dead_pix_analogue} for the dead pixel category, faulty pixels are rare, with only a small number of regions containing them. Regions with more than \SI{1}{\percent} faulty pixels are classified as failing this test, which corresponds to \SI{0.6}{\percent} of all tested regions.

\subsection{Pixel matrix performance}
\label{sec:pix_performance}

Threshold and noise are measured per pixel, while the fake-hit rate is measured per region (see Sec.~\ref{sec:threshold_noise_fhr}). Pixel matrix performance is then evaluated in terms of threshold and noise uniformity, and fake-hit rate after masking the noisiest pixels. Regions with more than \SI{1}{\percent} faulty pixels, either excessively noisy or with undetermined thresholds, are classified as having unsatisfactory performance, accounting for \SI{0.33}{\percent} of the tested regions.

\begin{figure}[!htb]
    \centering
    \includegraphics[width=1\linewidth]{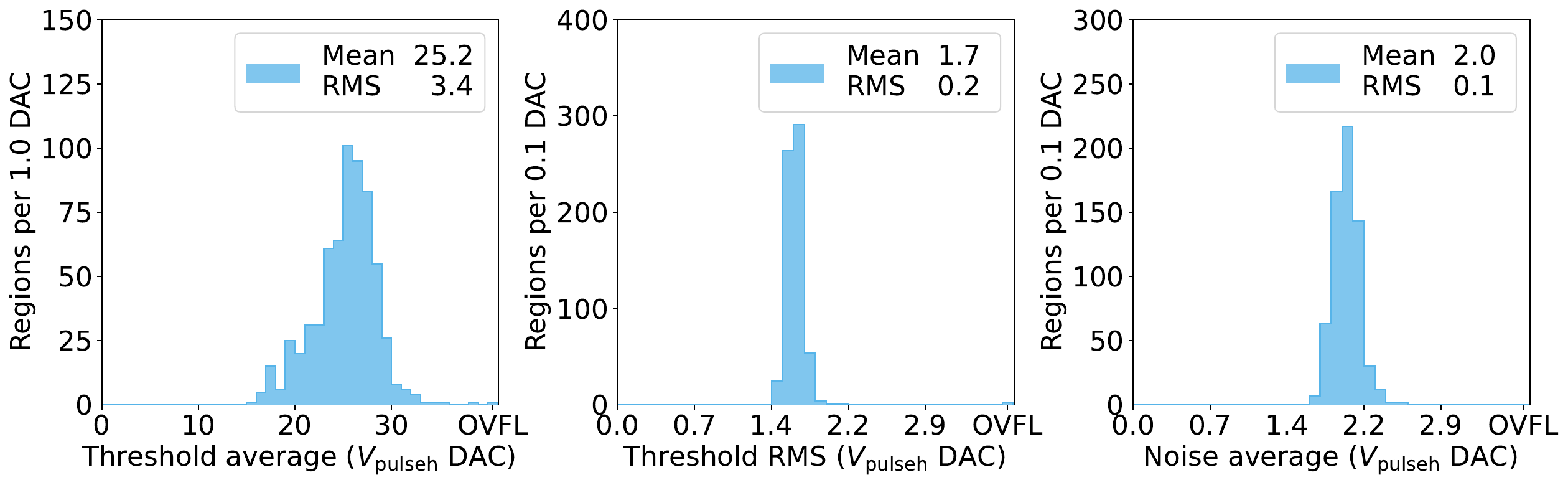}
    \caption{Distribution of the threshold average, threshold RMS, and noise average for all the tested regions with \SI{22.5}{\micro\metre} pixel pitch and standard front-end implementation. \num{1}~$\Vh$~DAC corresponds to about~\SI{8}{\ele}.}
    \label{fig:thr-noise-avg}
\end{figure}

Figure~\ref{fig:thr-noise-avg} shows the distributions of region-average threshold, threshold RMS, and average noise for all regions with \SI{22.5}{\micro\metre} pixel pitch and standard front-end implementation using nominal biasing settings (see Sec.~\ref{subsubsec:two_params_scan}). Values are reported in $\Vh$~DAC counts, as calibrating the injected charge at this scale (about 5000 regions in total, 642 shown here) was not feasible within the available time. The observed distribution spread of about \SIrange{10}{15}{\percent} reflects variations in pulsing capacitance (see Sec.~\ref{sec:pulsing_calib}), differences in analogue biasing (see Sec.~\ref{sec:analogue_biasing}), and variation in reverse bias voltage (discussed below).

\begin{figure}[!htb]
    \centering
    \includegraphics[width=1\linewidth] {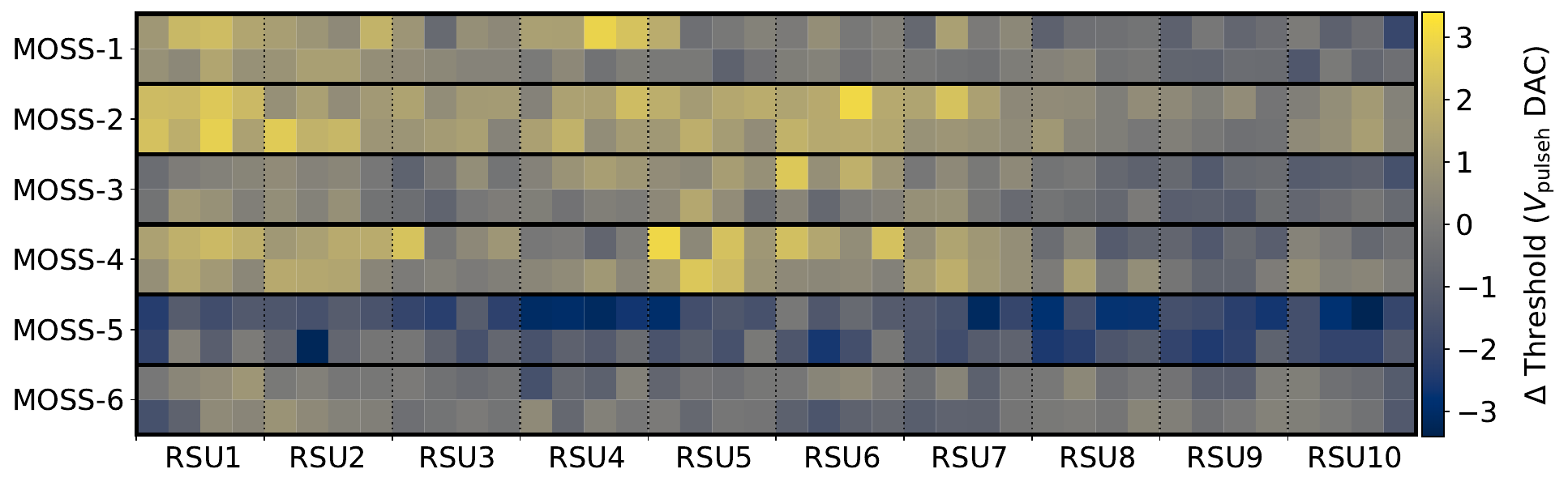}
    \caption{Average deviation from the mean region threshold as a function of wafer position. Two patterns are apparent, a horizontal gradient, and a lower average threshold in all regions corresponding to the MOSS-5 position. \num{1}~$\Vh$ DAC corresponds to about \SI{8}{\ele}.}
    \label{fig:thr-map}
 \end{figure}

Although the threshold can be set independently for each region, reducing the spread in Fig.~\ref{fig:thr-noise-avg} to a few percent, studying its variation across regions before any adjustment provides insight into potential design, manufacturing, or operational systematic effects. For example, the required operational margin for biasing parameters can be determined, and observed spatial patterns can be correlated with manufacturing steps or test setup effects.
Figure~\ref{fig:thr-map} shows a wafer-shaped map of average deviations from the mean region threshold across all wafers. The maximum average deviation per region is about \SI{25}{\ele}, or roughly \SI{13}{\percent}.
Two patterns are apparent, a horizontal gradient, and a lower average threshold in all regions corresponding to the position of MOSS number 5 on the wafer (see Fig.~\ref{fig:ER1_processed_MOSS}). The latter is linked to three specific sensors (out of fourteen) in the MOSS-5 position. Their lower average threshold is explained by the high substrate bias currents of \SIlist{26;44;53}{\milli\ampere} (see the PSUB distribution tail in Fig.~\ref{fig:endpoint_currents_power_on}). Higher currents reduce the substrate bias seen by the front-end, lowering the threshold.
The origin of the horizontal gradient remains unclear, though effects related to the test setup have been excluded.

\begin{figure}[hbt]
     \centering
     \begin{subfigure}[]{0.49\textwidth}
         \includegraphics[width=\textwidth]{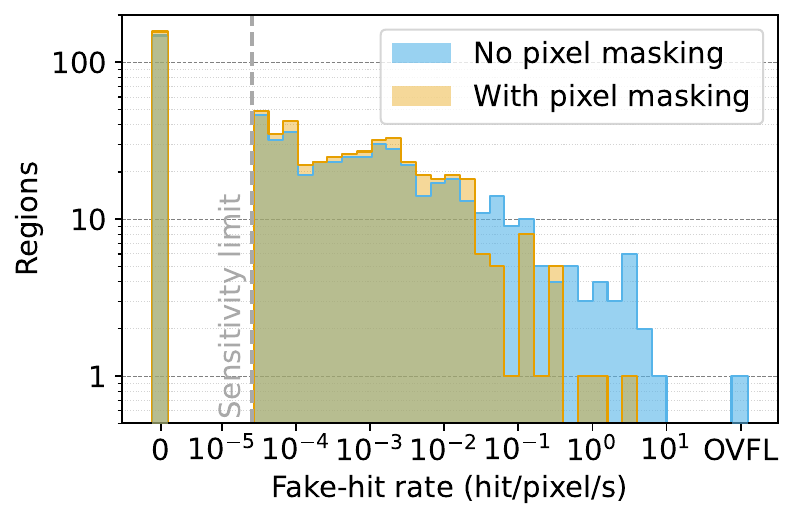}
         \caption{Fake-hit rate before and after masking.}
    \label{fig:fhr_w/o_mask}
     \end{subfigure}
\begin{subfigure}[]{0.49\textwidth}
         \includegraphics[width=\textwidth]{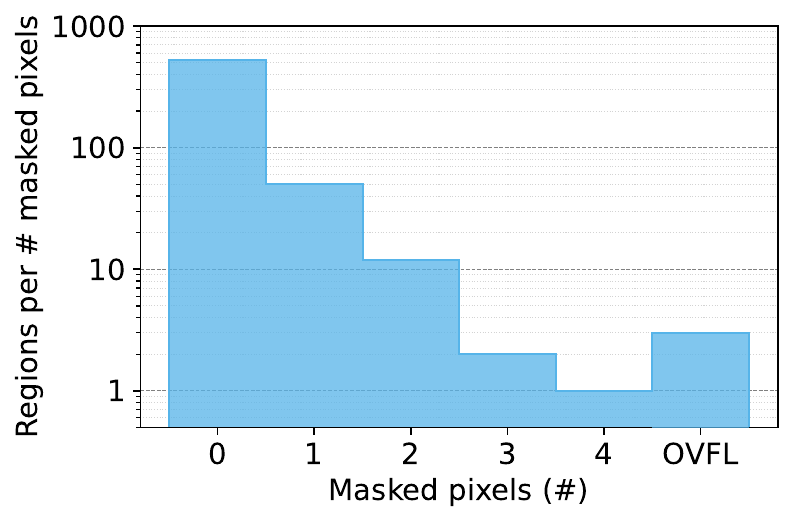}
        \caption{Number of masked pixels}
       \label{fig:fhr_masked_pixels}
     \end{subfigure}
\caption{Fake-hit rate performance of regions with \SI{22.5}{\micro\metre} pixel pitch and standard front-end variant. Masked pixels are those that occur in more than \SI{1}{\percent} of events (see Sec.~\ref{sec:threshold_noise_fhr}). The sensitivity limit is determined by the number of events, i.e. the duration of the measurement.}
\label{fig:fhr-dist}
\end{figure}

Figure~\ref{fig:fhr-dist} shows the fake-hit rate measurements for all regions with \SI{22.5}{\micro\metre} pixel pitch and standard front-end implementation (the same as in Fig.~\ref{fig:thr-noise-avg}). As discussed in Sec.~\ref{sec:threshold_noise_fhr}, noisy pixels are masked and excluded from the fake-hit rate calculation. In most cases, no noisy pixels are identified, and if present, masking a few is usually sufficient to significantly reduce the fake-hit rate. The spread in the fake-hit rate distribution corresponds to the spread of the threshold distribution (see Fig.~\ref{fig:thr-noise-avg}).

\subsection{Left End-Cap}
\label{sec:lec}

One of the development goals for the MOSS sensor was to prototype and validate the on-sensor data transfer with circuits extending over the full sensor length and crossing multiple stitching boundaries (see Sec.~\ref{subsec:architecture}) between the RSU and reaching the LEC.
Since future ITS3 sensors are expected to rely exclusively on readout via one of the extremities, it is important to validate the response of MOSS half-units when read out through the stitched communication backbone and the data interfaces on the LEC. To do so, several tests were repeated with half-units steered and read out via the LEC instead of their individual interfaces on the long edges. These tests inspect digital periphery functionalities, pixel matrix readout, and pixel matrix performance. Comparing the results between the two operating modes makes it possible to determine whether any issues arise from 1) the communication between the half-units and the LEC over the stitched communication backbone or 2) from the LEC itself.
Out of 78~sensors tested via the LEC, one issue for each type was identified, resulting in 21~additional regions (about \SI{0.7}{\percent}) not passing this check. In pixel matrix performance tests, threshold and fake-hit rate measurements were found to be consistent across both readout schemes.

\subsection{Half-unit and region yield}
\label{sec:yield}

\begin{figure}[!ht]
    \centering
    \includegraphics[width=0.6\linewidth]{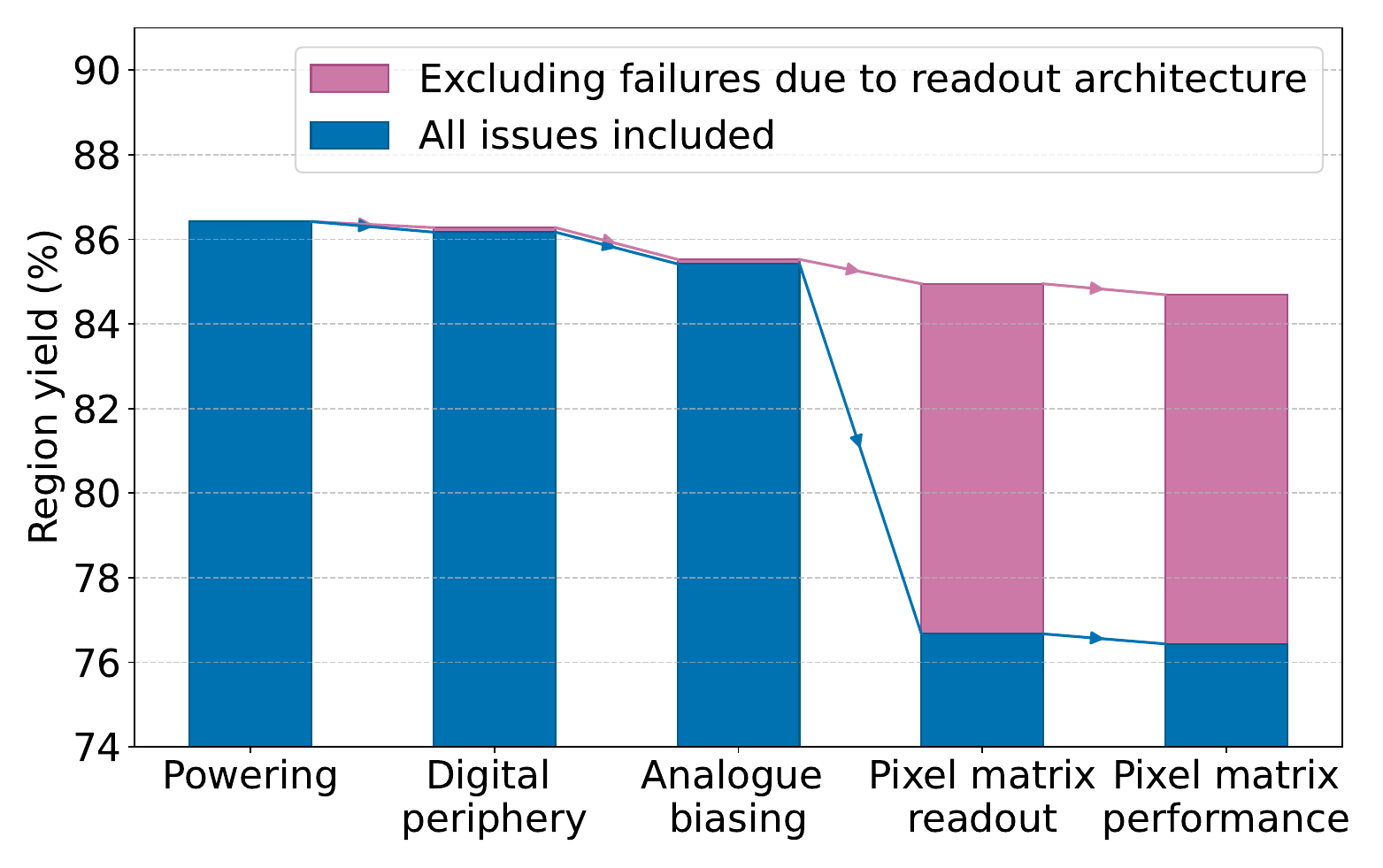}
    \caption{Yield losses, normalised per region, across the different sensor block verifications.}
    \label{fig:dropout}
\end{figure}

Figure~\ref{fig:dropout} summarises the yield losses, normalised per region, across the different testing phases described in the previous sections. The largest yield loss occurs during the powering test due to the faults in the power network, an issue which is not expected to affect forthcoming sensor designs (see Sec.~\ref{sec:powering}).
As discussed in Sec.~\ref{sec:matrix_readout}, yield loss in the matrix readout can result from issues in the simplistic readout architecture used in this prototype. Since these issues were anticipated in the sensor design, they can be excluded from the yield loss estimate.
All other tests indicate a robust design, with yield losses remaining below one percent at each stage. Overall, about \SI{76}{\percent} of regions pass the full test sequence considering all issues, and \SI{85}{\percent} when excluding the failures related to the readout-architecture limitations. 
Excluding also the powering issues, the region yield observed in functional tests is above \SI{98}{\percent}.

\begin{figure}[!hbt]
     \centering
     \includegraphics[width=1.0\textwidth]{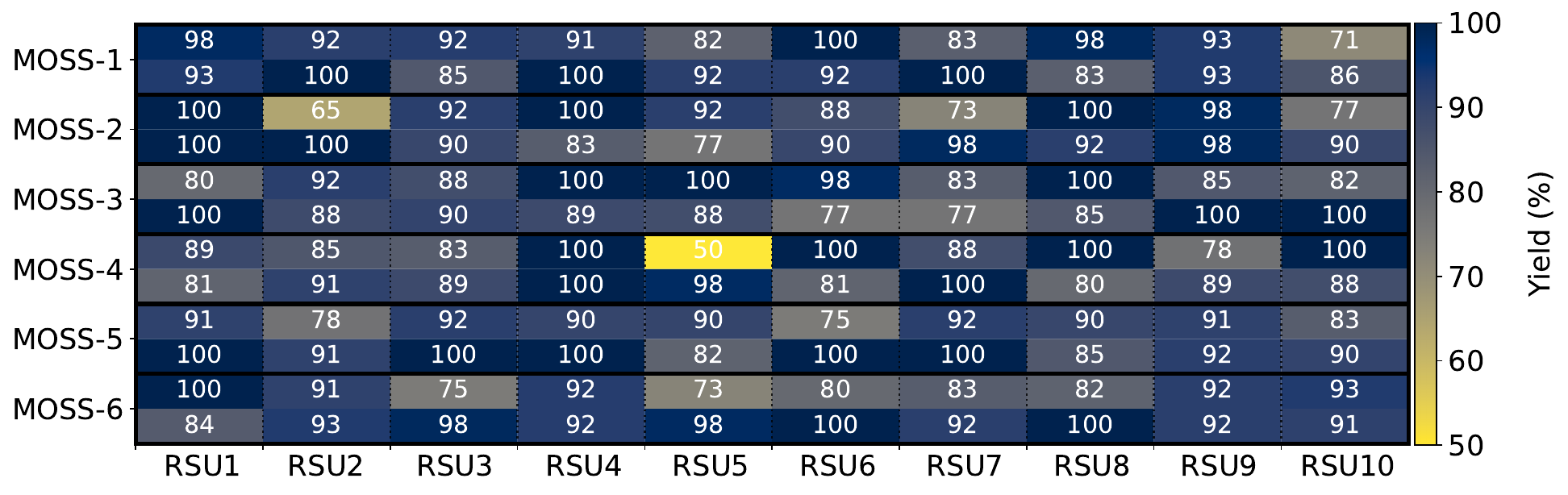}
    \caption{Half-unit yield as a function of wafer position. No specific patterns are apparent.}
    \label{fig:yield-map}
\end{figure}

Figure~\ref{fig:yield-map} shows the functional yield of half-units, including readout architecture issues, as a function of their wafer position. No specific patterns are apparent. Significantly, no top-bottom asymmetry within MOSS sensors is observed, indicating that the higher matrix-layout density of the bottom half-units due to the smaller pixel pitch did not cause any statistically significant yield loss.

\begin{figure}[!hbt]
    \centering
    \includegraphics[width=0.75\linewidth,]{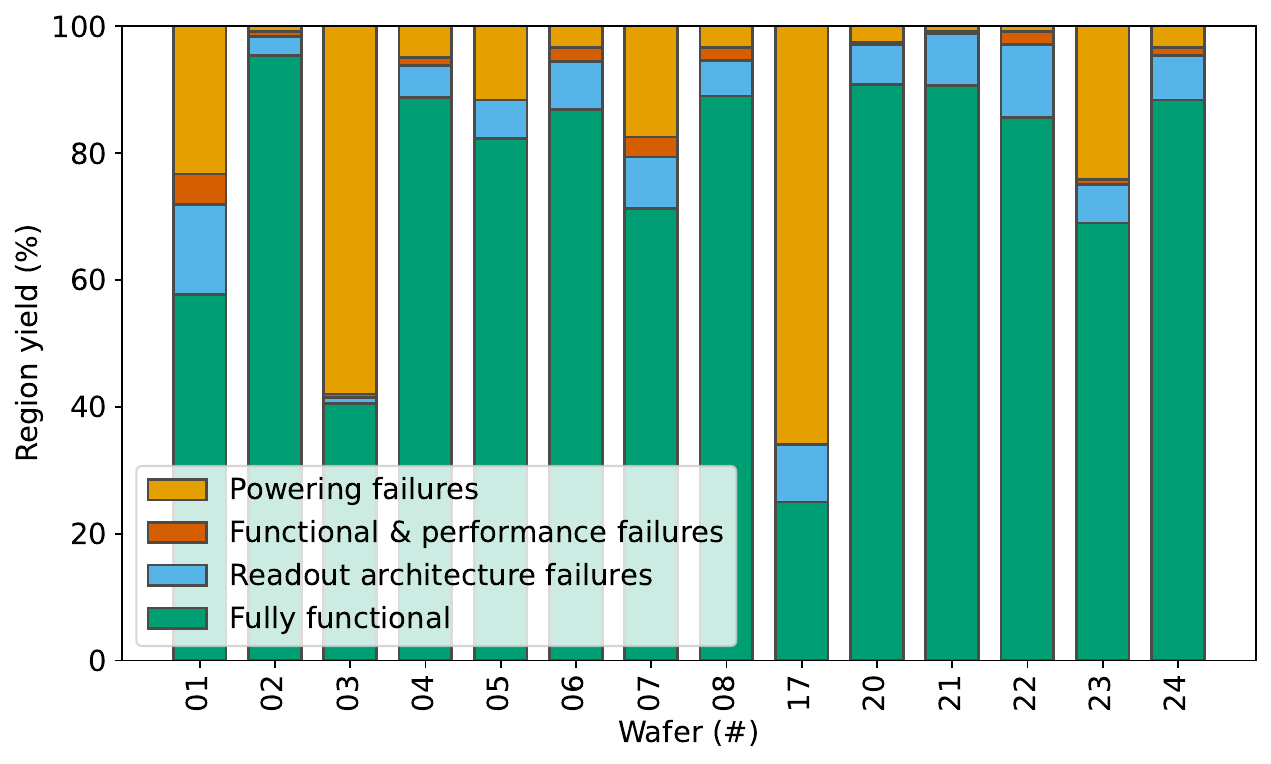}
    \caption{Yield overview per wafer. For each wafer, the fraction of regions in different categories is shown as a stacked representation. Powering yield loss is more common in odd-numbered wafers than in even-numbered wafers.}
    \label{fig:wafer-yield}
\end{figure}

The half-unit yield values per wafer are shown in Fig.~\ref{fig:wafer-yield}, for each of the 14 wafers systematically tested. The region yields are reported with \mbox{four categories}: failures of powering, functional but not meeting performance criteria, failures attributable to the sensor readout architecture, and fully functional.
Powering yield loss is more common in odd-numbered wafers than in even-numbered wafers. An analysis of processing control and monitoring data confirmed the existence of small differences between the odd-numbered and even-numbered wafers for some of the characteristics related to the metal interconnects. This correlates with the observed yield fluctuations and is attributed to the two wafer subsets being processed under slightly different conditions.

\subsection{Discussion on the yield of ITS3 layers}
\label{sec:layer_yield}

The series-testing data can be used to estimate the counts of ITS3 half-layers that could be produced in the hypothetical scenario of building them with MOSS sensors and with the yield observed within the ER1 engineering batch.
Each half-layer can be produced from a single wafer by dicing a set of adjacent MOSS sensors that meet the acceptance criteria.
The ITS3 half-layers for Layer 0, 1, and 2 would need respectively four, five or six adjacent functional MOSS sensors on one wafer. Each ER1 wafer contains six adjacent sensors (see Sec.~\ref{subsec:architecture}), allowing multiple possible mappings of half-layers to wafers.
The ITS3 requirements tolerate loosing up to \SI{2}{\percent} of the sensitive area~\cite{ITS3_TDR} due to local defects, which for MOSS corresponds to \SI{2}{\percent} region failures.
The following estimates of acceptable half-layer counts are based on 12~fully tested\footnote{\label{fn:layeryield}The remaining two wafers contained one mechanically broken sensor each, and thus were excluded from this exercise.} MOSS wafers, a half of a typical production lot, limiting the total number of possible combinations for each half-layer. Priority was given to achieving equal numbers of different half-layers rather than favouring a particular layer size.
Finally, it should be emphasized that these estimates are based on an exploratory prototype sensor, which still lacks features needed for ITS3, such as high-speed serialisers and the ability to power the entire sensor exclusively from the end-caps.

\begin{figure}[!ht]
    \centering
    \includegraphics[width=0.75\linewidth,trim=0 0.5cm 0 0.5cm,clip]{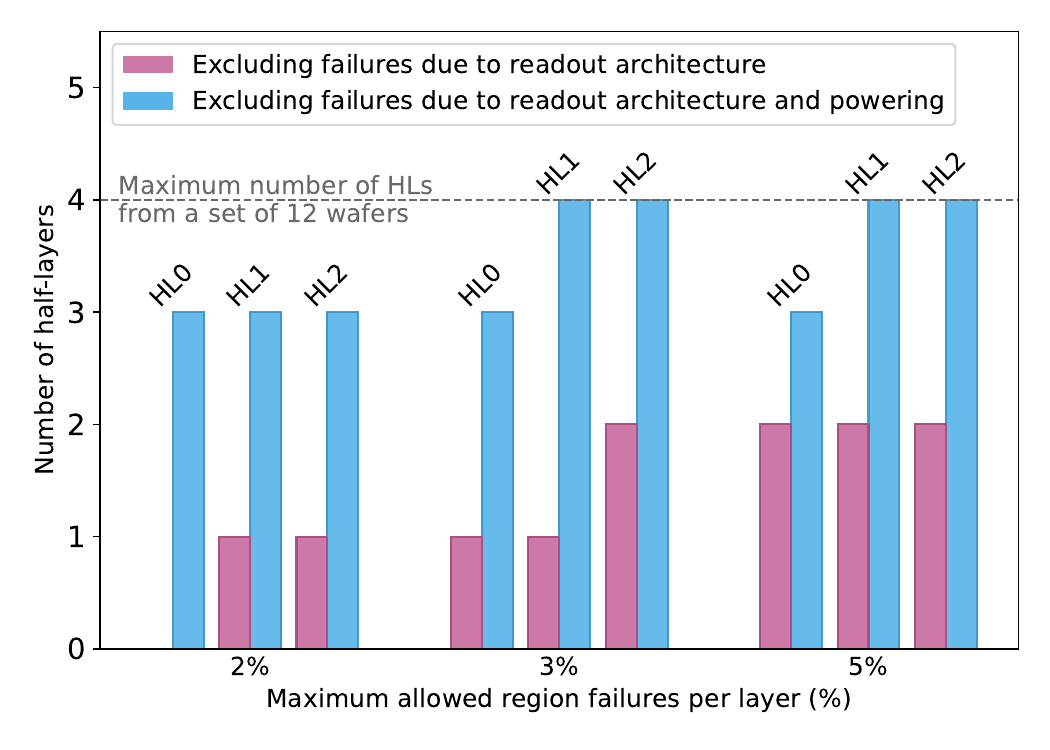}
    \caption{The number of ITS3 half-layers (HL) that can be successfully assembled with MOSS sensors from a set of 12 wafers\footref{fn:layeryield} under different assumptions on the allowed region failure rate and the types of excluded failures. Both scenarios are expected to be compatible with the final production.}
    \label{fig:wafer-yield-3}
\end{figure}

Figure~\ref{fig:wafer-yield-3} shows the number of ITS3 half-layers that can be successfully assembled from a set of twelve fully analysed wafers, under different acceptance assumptions. Two scenarios are presented, following the reasoning in the previous section: (a) excluding only region failures attributed to the readout architecture, and (b) excluding both readout architecture and powering-related failures.
The latter condition is less restrictive for sensor quality as it assumes that manufacturing imperfections associated with the metal stack can be fully mitigated in the future productions, which is yet to be demonstrated.
Three cases are considered, i) region failures amount to less than \SI{2}{\percent}, ii) less than \SI{3}{\percent}, and iii) less than \SI{5}{\percent}. If \SI{2}{\percent} of failures are tolerated than the scenario which excludes both readout architecture and powering-related failures would provide enough sensors to assemble the full ITS3 barrel, given that two half-layers per layer size are needed and three for each one are found. In the first scenario, which excludes only region failures attributed to the readout architecture, one would have to tolerate \SI{5}{\percent} of region failures to be able to obtain two half-layers of each type from the given twelve wafers. However, increasing the initial number of wafers would likely allow assembly of a full ITS3 barrel even under the most stringent conditions of less than \SI{2}{\percent} region failures.

\section{Single-Event Effect measurements} 
\label{sec:see_seu}

The collisions in the LHC create a radiation environment with a flux of particles that can generate stochastic Single-Event Effects (SEE), and in particular Single-Event Upset (SEU) and Single-Event Latch-up (SEL) in CMOS circuits. The future ITS3 sensor shall exhibit sufficient robustness against SEU to achieve reliable operation under these operating conditions. It shall also have low sensitivity to SEL to prevent potentially destructive radiation-induced over-currents.
Dedicated SEE tests were carried out on the MOSS sensor, which does not implement TMR redundancy techniques, to quantitatively evaluate the impact of Single Event Effects and to assess the potential need for mitigation measures in the final ITS3 sensor design.

Sensitivity to SEU was measured by programming specific data patterns into sensor registers and recording bit flips as a function of irradiation type, flux (typically hadrons), and exposure time. SEU characterisation tests were performed at the NPI cyclotron in Řež, using \SI{30}{\mega\eV} proton beams~\cite{npi}. Several MOSS registers implemented with flip-flops, performing various functions and located in different regions of the sensor, were simultaneously irradiated.
All monitored registers exhibited SEU cross sections between \SI{1e-14}{\square\cm\per\bit} and \SI{4e-14}{\square\cm\per\bit}, consistent with previous results from dedicated SEU test sensors fabricated in the same CMOS technology. 

While SEUs are generated directly by the products of the primary collisions, SEL events are caused indirectly by such particles through the production of nuclear-recoil fragments inside the silicon sensor. Such fragments can have a Linear Energy Transfer (LET) sufficiently high to trigger a SEL. On the other side the LET of these fragments has been shown to not exceed \SI{15}{\mega\eV\square\cm\per\milli\gram} under the operational conditions of the LHC~\cite{rgalia}. The sensitivity to SEL is assessed by detecting and measuring the frequency of latch-up events as a function of a flux of particles with a well defined LET. A latch-up event is characterised by a persistent increase of one of the sensor supply currents due to the creation of a parasitic, self-sustaining thyristor structure in the sensor. Resetting such structures requires power-cycling either the entire sensor or the smallest sub-sensor power domain where SEL has occurred. The sensitivity to SEL is highly dependent on specificities of the detailed layout of a given circuit 
and has therefore be tested on the MOSS sensor to identify locations eventually requiring modifications of the the circuit layout in the final sensor.

SEL tests were conducted at the Heavy Ion Facility of UCL (Louvain-la-Neuve, BE) and at the BASE facility of LBNL (Berkeley, US). Ion beams with LET between \SIrange{3.3}{62.5}{\MeV\square\cm\per\milli\gram} and fluxes between \SIrange{500}{15000}{ions\per\square\cm\per\second} were used, with about \SI{10}{\percent} homogeneity over an irradiation area with a diameter of \SI{2.5}{\cm}.
To avoid beam degradation, the setup was installed in a vacuum vessel, and the sensor was connected to a cooling circuit for temperature stabilisation. During irradiation, the supply currents and the correct functioning of the sensor were continuously monitored. Collimators and movable stages were also used to expose selectively small areas of the sensors to better identify locations with high SEL sensitivity.
The threshold for the detection of over-currents was set to \SI{50}{mA} to enable clear identification of typical latch-up current pulses.

\begin{figure}[!htb]
    \centering
    \includegraphics[width=0.75\linewidth,trim=0 0.4cm 0 0.4cm,clip]{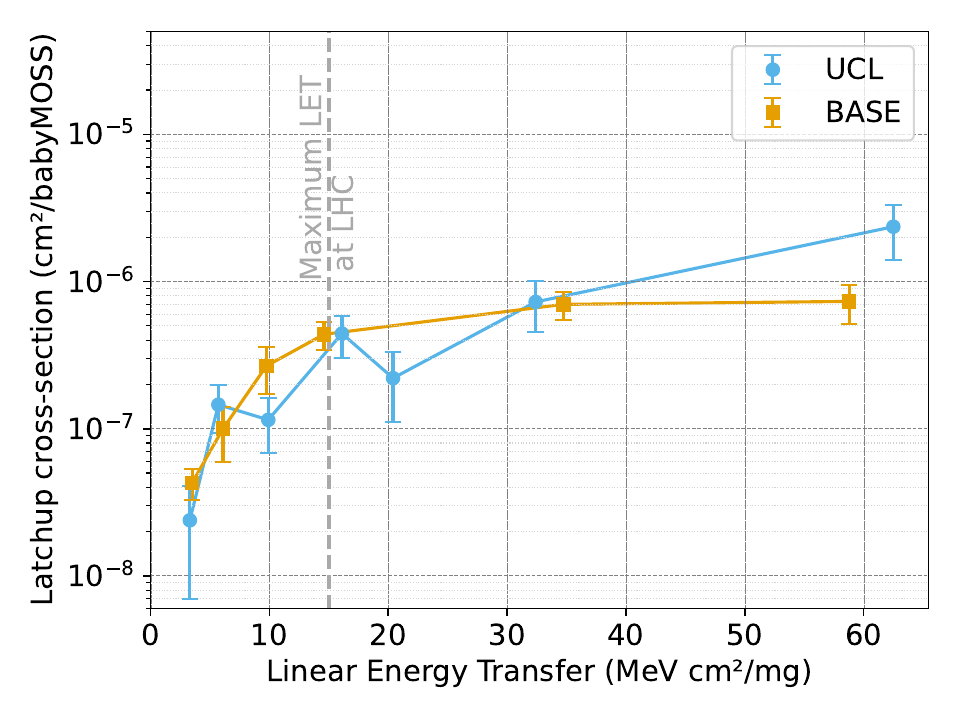}
    \caption{SEL cross section per total area irradiated babyMOSS as a function of LET measured at the Heavy Ion Facility of UCL (Louvain-la-Neuve, BE) and at the BASE facility of LBNL (Berkeley, US). The dashed line indicates the maximum LET that fragments generated in silicon can reach under LHC operational conditions.}
    \label{fig:bmSEL}
\end{figure}

Figure~\ref{fig:bmSEL} presents the measured SEL cross-section for a babyMOSS device irradiated without any collimators as a function of the LET of the incident ion species. The data exhibit the characteristic steep rise in cross section at low LET values, followed by a more gradual increase for LET values above approximately \SIrange{20}{30}{\mega\eV\square\cm\per\milli\gram}.
Despite the relatively high detection threshold of \SI{50}{\milli\ampere}, SEL events were observed at LET values below \SI{15}{\mega\eV\square\cm\per\milli\gram}. Targeted irradiations using various collimators enabled the localization of SEL-sensitive regions to the periphery of the sensor. Notably, no SEL events were detected when irradiating the pixel array.
An insufficient density of well contacts was identified in certain peripheral components, prompting corrections for subsequent iterations of the sensor design. SEL-sensitivity testing will be repeated on future designs to assess the likelihood of latch-up under the expected operational conditions of ITS3.

\section{Conclusions} 
\label{sec:conclusions}

This paper presents an overview of the design, as well as the results of a comprehensive testing and characterisation campaign, of the MOSS sensor, a novel wafer-scale prototype CMOS MAPS sensor, assembled from ten identical sensors using sophisticated stitching techniques.
Its design and testing represented key milestones in the development of stitched MAPS and the prototyping of future ALICE ITS3 sensors, while also generating valuable know-how that will be reflected in the subsequent sensor design iteration.
The sensor is fully functional, and demonstrates the viability of implementing the stitching technique as a means to manufacture sensing devices of different size with one modular design, in line with the integration requirements of the ITS3 half-layers. 
Complex power distribution with multiple power nets, metal interconnects across stitching boundaries, and transmission of signals on-sensor over about \SI{25}{\cm} have been validated.

A total of 82~full MOSS sensors were tested systematically in the lab, each one containing 10~Repeated Sensor Units, amounting to 6560~powered and controlled pixel matrices, and over half a billion individually characterised pixels.
The testing of the ER1 lot identified an excess rate of faults in the sensor power grid.
A detailed failure analysis was conducted and enabled the identification of manufacturing imperfections associated with the metal stack, newly introduced by the foundry to meet the specific requirements of the project. Feedback was shared with the foundry, enabling the identification of the root cause and the implementation of corrective measures that are expected to drastically reduce the likelihood of these defects in future productions with the same metal stack~\cite{metal_stack_paper}.

Despite these faults, the qualification demonstrated correct operation and acceptable performance of the individual building blocks and a resilient design, with a comprehensive functional yield exceeding \SI{98}{\percent}.
Extrapolations based on the systematic functional qualification of the MOSS sensors show a yield sufficient to construct ITS3 half-layers from single wafers.
The results of the systematic laboratory analysis did not find significant variations of characteristics inside a wafer and from wafer-to-wafer, once the faults related to the metal-stack issue were excluded.
Significantly, no change of the functional failure rates was observed in relation to the different layout densities prototyped in the two halves of the MOSS sensor, providing experimental evidence of the possibility to relax some of the conservative margins adopted in the layout of the pixel array with larger pitch.

A subset of sensors underwent extensive pixel matrix characterisation in the laboratory and with ionising-particle beams. The required performance, a detection efficiency above \SI{99}{\percent} and a fake-hit rate below \SI{0.1}{hits\per\pixel\per\second} was achieved.
Spatial resolution was found to be \SIrange{5}{5.5}{\micro\metre} for \num{22.5}-\si{\micro\metre}-pitch pixels and \SIrange{4}{4.5}{\micro\metre} for \num{18}-\si{\micro\metre}-pitch pixels, indicating that an intermediate pixel pitch will be sufficient to meet the ITS3 target of \SI{5}{\micro\metre}.
No significant differences of key performance figures were observed across the pixel variants prototyped in MOSS.
Cross-coupling effects between digital signals and sensitive analogue nodes were identified and thoroughly studied, providing useful knowledge for mitigation in future design iterations.
Measurements of SEL cross sections with heavy-ion beams revealed a sensitivity of MOSS to single-event latch-ups for LET below \SI{15}{\mega\eV\square\cm\per\milli\gram}, localised in specific digital peripheral blocks and attributed to sub-optimal contacting of wells. 

In conclusion, the MOSS sensor demonstrates the viability of stitching to implement large area MAPS and exhibits promising performance and yield for the ALICE ITS3 upgrade.
Remaining challenges for the next sensor include demonstrating high-speed data transmission, both on- and off-sensor, and managing the voltage drops resulting from power being supplied solely at the short sensor edges -- features that are absent in the MOSS design.
The comprehensive characterisation of MOSS provides valuable insights into the sensor's behaviour under different operating conditions and irradiation levels, paving the way for the design of the final ITS3 sensor and for further developments and applications.

\newenvironment{acknowledgement}{\relax}{\relax}
\begin{acknowledgement}
\section*{Acknowledgements}
Filip K\v{r}\'i\v{z}ek acknowledges the support by the Ministry of Education, Youth and Sports of the Czech Republic project LM2023040.

Zijun Zhao acknowledges the support by the National Key Research and Development Program of China (2022YFA1602103).

Tommaso Fagotto and Giovanni Vecil acknowledge the support by the Italian Ministry of Foreign Affairs and International Cooperation (project PGR12370).

This research has received funding support from NSRF via the Program Management Unit for Human
Resources \& Institutional Development, Research and Innovation [B46G680100], and NSTDA under
the contract number JRA-CO-2563-14066-TH.

The authors thank the Optimato team at the Elettra Sincrotrone for providing their facility and expert support in conducting the X-ray fluorescence measurements.
\end{acknowledgement}

\bibliography{references}

\end{document}